\documentclass[]{article}

\usepackage{amsfonts,amsmath,amssymb,amsthm,mathrsfs}
\usepackage[numbers,sort]{natbib}
\usepackage[utf8]{inputenc}
\usepackage[english]{babel}
\usepackage{bbm}
\usepackage{enumitem}
\usepackage{color}
\usepackage{comment}
\usepackage[margin=1.045in]{geometry}
\usepackage{setspace}
\usepackage{scalerel}
\usepackage{lipsum}
\usepackage{float}
\usepackage{caption}
\usepackage{subcaption}
\usepackage{parskip}
\usepackage[bottom]{footmisc}
\usepackage[toc, page]{appendix} 
\usepackage{color}

\usepackage[colorlinks=true,linkcolor=blue,citecolor=blue,urlcolor=black,pdfborder={0 0 0}]{hyperref}
\usepackage{cleveref}

\crefname{theorem}{Theorem}{Theorems}
\crefname{thm}{Theorem}{Theorems}
\crefname{lemma}{Lemma}{Lemmas}
\crefname{lem}{Lemma}{Lemmas}
\crefname{remark}{Remark}{Remarks}
\crefname{prop}{Proposition}{Propositions}
\crefname{defn}{Definition}{Definitions}
\crefname{corollary}{Corollary}{Corollaries}
\crefname{conjecture}{Conjecture}{Conjectures}
\crefname{question}{Question}{Questions}
\crefname{chapter}{Chapter}{Chapters}
\crefname{section}{Section}{Sections}
\crefname{figure}{Figure}{Figures}

\newtheorem{theorem}{Theorem}

\newtheorem{proposition}[theorem]{Proposition}

\theoremstyle{remark}

\numberwithin{theorem}{section}

\newcommand{\figuretag}[1]{%
  \addtocounter{figure}{-1}%
  \renewcommand{\thefigure}{#1}%
}

\title{Economics and Optimal Investment Policies of Attackers and Defenders in Cybersecurity}

\author{Austin Ebel$^*$ \and Debasis Mitra\thanks{Department of Electrical Engineering, Columbia University. Email: \{abe2122,\hspace{0.2em}debasismitra\}@columbia.edu}}

\begin{document}

\maketitle

\begin{abstract}
In our time cybersecurity has grown to be a topic of massive proportion at the national and enterprise levels. Our thesis is that the economic perspective and investment decision-making are vital factors in determining the outcome of the struggle. To build our economic framework, we borrow from the pioneering work of Gordon and Loeb in which the Defender optimally trades-off investments for lower likelihood of its system breach. Our two-sided model additionally has an Attacker, assumed to be rational and also guided by economic considerations in its decision-making, to which the Defender responds. Our model is a simplified adaptation of a model proposed during the Cold War for weapons deployment in the US. Our model may also be viewed as a Stackelberg game and, from an analytic perspective, as a Max-Min problem, the analysis of which is known to have to contend with discontinuous behavior. The complexity of our simple model is rooted in its inherent nonlinearity and, more consequentially, non-convexity of the objective function in the optimization. The possibilities of the Attacker's actions add substantially to the risk to the Defender, and the Defender's rational, risk-neutral optimal investments in general substantially exceed the optimal investments predicted by the one-sided Gordon-Loeb model. We obtain a succinct set of three decision types that categorize all of the Defender’s optimal investment decisions. Also, the Defender's optimal decisions exhibit discontinuous behavior as the initial vulnerability of its system is varied. The analysis is supplemented by extensive numerical illustrations. The results from our model open several major avenues for future work.
\end{abstract}

\section{Introduction}

In recent times the role of cybersecurity has grown at a pace comparable to that of information technology in society. This is to be expected since its failure has the potential to inflict damage wherever IT is extensively deployed, from national security to transportation, health care, banking, and energy. The number and variety of attackers, which range from state actors and terrorists to criminals and hackers, has grown similarly, as have their resources \cite{ref:Clark}. Just at the level of commercial enterprises, the projected current annual loss from cybercrime is estimated to be \$945 billion, almost double the corresponding amount of \$500 billion in 2018 \cite{ref:Smith}. IBM has estimated the cost of a (commercial) data breach in 2021 to be \$4.24 million, up 10\% from the previous year \cite{ref:IBM}.

Cybersecurity defense and investments have kept pace as well. For example, at the national level in the USA, consider the linked combination of the National Vulnerability Database (NVD), the Common Vulnerabilities and Exposures (CVE) list, and the NIST Common Vulnerability Scoring System (CVSS). NVD is the NIST-maintained database of vulnerability management data, which includes security checklist references, security-related software flaws, misconfigurations, product names, and impact metrics \cite{ref:NVD}. CVE is a list of publicly disclosed vulnerabilities and exposures that is maintained by MITRE, and it is fully synchronized with NVD. CVSS is the overall score assigned to a vulnerability, and the NVD may be used to find assigned CVSS scores \cite{ref:Mell}. This is indicative of the magnitude of investments in cybersecurity defense being made in the USA.

Our work in cybersecurity takes the economic perspective, and it examines the many facets of investment decision-making in the context of a model, which is simple to describe, and yet is of broad interest and rich in analytic and computational complexity. The model is an adaptation of a simplification of a model proposed during the Cold War for weapons deployment in the US \cite{ref:Danskin}. An essential property of the model is that it is two-sided, i.e., the Attacker and the Defender actively interact through the instrument of investments to further their respective goals of reaping gains from breaching the system, and minimizing expected losses by mitigating vulnerabilities. Both the Attacker and the Defender are assumed to be rational, risk-neutral and guided by economic consideration in their decision-making.

We build our economic framework on the important, pioneering work of Gordon and Loeb \cite{ref:Gordon} based on a one-sided model in which the Defender optimally invests in vulnerability mitigation. The Defender’s investment decision is based on a known nonlinear function, $S(z,v)$, which gives the probability of a breach in its system from attacks, and which depends on the system's initial vulnerability ($v$), and the subsequent investment ($z$). For two specific classes of the function $S(z,v)$, they also prove the striking result that the optimum Defender’s investment does not exceed $1/e$, i.e., about 37\%, of the expected loss from a breach. Subsequent work \cite{ref:Baryshnikov, ref:Lelarge} proved that the $1/e$ rule holds whenever $S(z,v)$ is log-convex in $z$. In our work we also assume log convexity in $z$ of the Defender’s function $S(z,v)$, and, in fact, one of our goals is to obtain an understanding of the optimum investment in the two-sided model for broad classes of such log-convex functions.

Our two-sided model may be viewed as a Stackelberg leader-follower game with the Defender as leader and the Attacker as follower \cite{ref:Gibbons}. Our model and analysis may also be viewed to be Max-Min. The closest related work is Danskin \cite{ref:Danskin}; however, our results here are quite different. Our goal is to extract a qualitative understanding of the quantitative processes in the optimization of investments. Our main obstacle is the provable nonconvexity of the Defender’s objective function\footnote{It is worth recalling the view of an eminent researcher, R.T. Rockafeller: ``The great watershed in optimization isn’t between linearity and nonlinearity, but convexity and nonconvexity." \cite{ref:Rockafellar}}, which puts the problem nominally in the intractable category. However, there is enough structure that we are able to exploit to show that there are only three distinct types of optimum investment. Building on this result, we next look at the point of transitions in the Defender’s optimum decision types as the initial vulnerability, $v$, is varied. We show that the points at which transitions occur are solutions of Fixed Point equations. While the total number of transitions are small, decisions across transitions are typically sharply different. The main features of the Defender’s optimal policy are traceable to the aforementioned nonconvexity. We also find that, in general, the investments predicted by the one-sided model of Gordon-Loeb \cite{ref:Gordon} have qualitatively different characteristics, and also underestimate the investments necessary to prudently defend the system against attackers.

Throughout the paper we provide visualizations of the analytic development, and illustrations of behavior identified in the analysis.

\subsection*{Other Related Work}

Anderson \cite{ref:Anderson1} gives an economic perspective of the many, often conflicting, factors that make realizing cybersecurity hard. For results from a detailed study of the costs of cybercrime see \cite{ref:Anderson2}. The relationships between the average cost of an attack and the expected gain to the Attacker, and similar considerations for the Defender, are examined in \cite{ref:Tiwari}. Data from Japan’s local government bodies, i.e., municipalities, on their investments in information security in relation to vulnerabilities are analyzed, and significant correlations are detected \cite{ref:Tanaka}.  

A broad overview from the systems perspective of procedures employed by Attackers and Defenders of information systems is in \cite{ref:Mazurczyk}. Our paper makes no attempt to analyze cybersecurity for networked systems, but it may provide a springboard for future work, in which case attack graph representations would be a natural framework, see \cite{ref:Wang}. Moving Target Defense is a technique in cyberdefense wherein the Defender reconfigures its system periodically, thus forcing the Attacker in each iteration to expend resources from a near cold start, see the review article \cite{ref:Lei}. The model and analysis of our paper may prove useful.

Game-theoretic analyses of network security is an area with a rich literature. It includes Sequential Stackelberg Games \cite{ref:Breton}, a special case of which corresponds to our model. In \cite{ref:Gueye} the system analyzed is a network of flows, in which the Attacker aims to disrupt selected flows, and the results yield a vulnerability metric.

The paper is organized as follows. In Section \hyperref[sect2]{2}, we introduce the basic variables and functions, and list our assumptions. The section introduces the two classes of functions $S(z,v)$ defined by Gordon and Loeb \cite{ref:Gordon}, which are used extensively in the paper to illustrate developments; also introduced is a function $\gamma(s)$ that parametrizes log convexity of $S(z,v)$ and is important in the analysis. The basic model describing the interaction of Attackers and Defenders is given in Sec. \hyperref[sect3]{3}. Section \hyperref[sect4]{4} formulates the Attacker’s problem and gives the solution. An analogy to the arms race during the Cold War is drawn. Sec. \hyperref[sect5]{5} tackles the Defender’s problem, and the shape of the gradient of the Defender’s objective function is determined. The solution to the Defender’s problem is given, and it is shown that there are three basic decision types. Sec. \hyperref[sect6]{6} gives the Fixed Point equations that define points of transition in the Defender’s decision types and results for certain classes. Various examples and illustrations of the results for the model are given. Sec. \hyperref[sect7]{7} concludes with a brief discussion of directions of possible future work.

\section{Preliminaries}
\label{sect2}

\subsection{Basic Variables, Functions, and Properties}
\label{sect:system_breach}

We define the following variables related to the Defender's system:

\begin{enumerate}[itemsep=-3px]
    \item $s = $ system breach probability, also referred to as (system) vulnerability
    \item $v =$ initial breach probability, also referred to as initial (system) vulnerability
    \item $z = $ incremental ``effort" by system Defender to mitigate system vulnerability. We distinguish effort from its financial cost to the Defender, with the latter given by $d\hspace{0.1em} z$, where $d$ is the Defender's unit cost of effort.
    \item $s = S(z,v)$, $S(.,.)$ is the system breach probability function
\end{enumerate}

\vspace{5px}

We make the following assumptions on the properties of the system breach probability function. Gordon-Loeb \cite{ref:Gordon} also assume \ref{itm:A1}--\ref{itm:A5}, whereas \ref{itm:A6} is new.

\begin{enumerate}[label=A\arabic*,itemsep=-3px]
    \item \label{itm:A1} \hspace{-4px}. $S(z,0)= 0, \ \forall z \geq 0$
    \item \label{itm:A2} \hspace{-4px}. $S(0,v) = v$
    \item \label{itm:A3} \hspace{-4px}. $S_z(z,v) = \frac{\partial S(z,v)}{\partial z} < 0, \ \forall z$ and $\forall v \in (0,1)$
    \item \label{itm:A4} \hspace{-4px}. $S_{zz}(z,v) > 0, \ \forall v \in (0,1)$
    \item \label{itm:A5} \hspace{-4px}. $S(z,v) \rightarrow 0$ as $z \rightarrow \infty, \ \forall v \in (0,1)$
    \item \label{itm:A6} \hspace{-4px}. $S_v(z,v) > 0, \ \forall z$ and $\forall v \in (0,1)$, i.e., system vulnerability increases with initial vulnerability
\end{enumerate}

\subsection{Effort Function}
\label{sect:effort_function}

We define the effort function $Z(.,.)$, where $z=Z(s,v)$. The function $Z$ is the inverse of the function $S$, i.e., $z \equiv Z(S(z,v),v), \ \forall z \geq 0$. Our analysis, perhaps unexpectedly, is primarily concerned with the behavior of the effort function, and not the system breach probability function. (Gordon-Loeb's analysis is focused on the function $S$).

\subsubsection*{Examples}
Gordon-Loeb \cite{ref:Gordon} define two classes of system breach probability functions, I and II. We have found these functions to be very useful in illustrating our analysis. We refer to these function classes as GL Class I and GL Class II.

\begin{enumerate}[label=(\roman*),itemsep=0ex]
    \item GL Class I. $S(z,v)=\frac{v}{(\alpha z + 1)^\beta}$, $\alpha > 0, \  \beta \geq 1$; $Z(s,v)=\frac{1}{\alpha} \left( \frac{v}{s}\right)^{1/\beta} - \frac{1}{\alpha}$
    \item GL Class II. $S(z,v)=v^{\alpha z + 1}$, $\alpha > 0$; $Z(s,v)= \frac{1}{\alpha} \frac{\log s}{\log v} - \frac{1}{\alpha}$ \ (natural log)
\end{enumerate}
\vspace{8px}
Since $\frac{\partial Z(s,v)}{\partial s} = Z_s(s,v) = \frac{1}{S_z(z,v)}$, hence from \ref{itm:A3},
\begin{equation}\tag{2.3}
\label{eq:2.3}
Z_s(s,v) < 0
\end{equation}
We will interpret $-Z_s(s,v) = \frac{\partial Z}{\partial (-s)}$ as the (positive) \textit{marginal effort for invulnerability} (for fixed initial vulnerability). Analogously,
\begin{equation}\tag{2.4}
\label{eq:2.4}
Z_v(s,v) = -\frac{S_v(z,v)}{S_z(z,v)} > 0
\end{equation}
is the marginal effort for maintaining invulnerability with increased initial vulnerability, with positivity following from \ref{itm:A3} and \ref{itm:A6}. Similarly,
\begin{equation}\tag{2.5}
\label{eq:2.5}
Z_{ss}(s,v) = -\frac{S_{zz}(z,v)}{\left\{ S_z(s,v)\right\}^3} > 0
\end{equation}

Hence, $Z(s,v)$ is a convex, decreasing function of $s$. Since $Z_{ss}(s,v) = \frac{\partial}{\partial (-s)}\big\{ \frac{\partial Z}{\partial (-s)} \big\}$, (\ref{eq:2.5}) implies that the marginal effort for invulnerability increases with increased invulnerability.

We make the additional assumption,
\begin{enumerate}[label=A\arabic*,itemsep=-3px,ref=2.6]
\setcounter{enumi}{6}
\item \label{eq:2.6} \hspace{-4px}. \hspace{0px}
$\begin{aligned}[t]
    Z_{sv}(s,v) < 0 \hspace{0.1em},
\end{aligned}$ \hspace*{\fill}(2.6)
\end{enumerate}

i.e., the marginal effort for invulnerability increases with increased initial vulnerability. We call this property ``effort complementarity", and comment on it below. Furthermore, we assume the following factored form of $Z_s(s,v)$:
\begin{enumerate}[label=A\arabic*,itemsep=-3px,ref=2.7]
\setcounter{enumi}{7}
\item \label{eq:2.7} \hspace{-4px}. \hspace{0px}
$\begin{aligned}[t]
    Z_s(s,v) = -\frac{f(v)}{g(s)} \hspace{0.1em}
\end{aligned}$ \hspace*{\fill}(2.7)
\end{enumerate}

\vspace{3px}

where, following (\ref{eq:2.3}), $f(v)$ and $g(s)$ are positive $\forall v,s \in [0,1]$, and are unique to within constants of proportionality. With this form, it follows from (\ref{eq:2.6}) and (\ref{eq:2.5}) that,
\begin{equation}\tag{2.8}
\label{eq:2.8}
f_v(v) > 0 \quad \text{and} \quad g_s(s) > 0
\end{equation}
We verify that the GL functions satisfy the above assumptions and the factored form.
\begin{enumerate}[label=(\roman*),itemsep=0ex]
    \item GL Class I. 
    \vspace{-10px}
    \begin{align}
        &Z_s(s,v) = -\frac{v^{1/\beta}}{\alpha \beta} \frac{1}{s^{(\beta+1)/\beta}} \tag{2.9} \\[3px]
        &f(v) = v^{1/\beta}; \quad g(s) = \alpha \beta s^{(\beta+1)/\beta} \label{eq:2.10} \tag{2.10} \\[3px]
        &Z_{sv}(s,v) = -\frac{1}{\alpha \beta^2} \frac{1}{v^{(\beta-1)/\beta}} \frac{1}{s^{(\beta+1)/\beta}} \tag{2.11}
    \end{align}
    \item GL Class II.
    \vspace{-10px}
    \begin{align}
        &Z_s(s,v) = \frac{1}{\alpha (\log v) s} \tag{2.12}\\[3px]
        &f(v) =-\frac{1}{\log v}; \quad g(s) = \alpha s \label{eq:2.13} \tag{2.13}\\
        &Z_{sv}(s,v) = -\frac{1}{\alpha \left( \log ^2 v\right) v s} \tag{2.14}
    \end{align}
\end{enumerate}

\subsection{Log Convexity}

We assume that $S(z,v)$ is log-convex in $z$ for all fixed $v$, i.e., $\log S(z,v)$ is convex, which translates to,
\begin{equation}\tag{2.15}
\label{eq:2.15}
\frac{S_{zz}(z,v) \hspace{0.1em} S(z,v)}{\left\{ S_z(z,v)\right\}^2} \geq 1
\end{equation}

It is useful to parameterize the log convexity property by the (function) $\gamma(z,v)$:
\begin{equation}\tag{2.16}
\label{eq:2.16}
    \gamma(z,v) = \frac{S_{zz}(z,v) \hspace{0.1em}  S(z,v)}{\left\{ S_z(z,v)\right\}^2} - 1
\end{equation}

Hence, log convexity holds if $\gamma(z,v) \geq 0$. Making use of the factored form in (\ref{eq:2.7}), and noting that,
\begin{equation*}
    \frac{S_{zz}(z,v) \hspace{0.1em} S(z,v)}{S^2_z(z,v)} = \frac{s \hspace{0.1em} g_s(s)}{g(s)}
\end{equation*}
it follows that,
\begin{equation}\tag{2.17}
\label{eq:2.17}
    \gamma(s,v) = \frac{s \hspace{0.1em} g_s(s)}{g(s)} - 1
\end{equation}
Henceforth we abbreviate $\gamma(s,v)$ to $\gamma(s)$; it has an important role in our analysis.

It was shown independently by Baryshnikov \cite{ref:Baryshnikov} and Lelarge \cite{ref:Lelarge} that Gordon and Loeb's celebrated $1/e$ rule \cite{ref:Gordon} holds if the system breach probability function, $S(z,v)$, is log-convex, and also that both Class I and II functions used by Gordon and Loeb to demonstrate the rule are log-convex.

With regard to the previously noted GL Class I and GL Class II system breach probability functions, note the following:
\begin{enumerate}[label=(\roman*),itemsep=0ex]
    \item GL Class I. $\gamma(s) \equiv 1/\beta$, a constant. Gordon-Loeb \cite{ref:Gordon} require $\beta \geq 1$, so that $\gamma \leq 1$. In general, we will not place this restriction. However, there are some significant qualitative differences that appear in the analysis depending on whether $\gamma \leq 1$ or $\gamma > 1$. Hence, when the need arises to assume $\gamma \leq 1$, we will specify ``GL Class I with $\gamma \leq 1$", and similarly for $\gamma > 1$, it being understood that otherwise $\gamma$ is any nonnegative constant in GL Class I.
    \item GL Class II. $\gamma(s) \equiv 0$.
\end{enumerate}

For $g(s)=O(s^\delta)$ as $s \rightarrow 0$, it may be verified from (\ref{eq:2.17}) that,
\begin{equation} \tag{2.18}
\label{eq:2.18}
    \gamma(s) \sim \delta - 1.
\end{equation}
Hence,
\begin{equation} \tag{2.19}
\label{eq:2.19}
    \delta > (=) \ (<) \ 2 \ \ \text{if} \ \ \gamma(0) > (=) \ (<) \ 1.
\end{equation}
The above dependence of the asymptotic behavior of $g(s)$ as $s \rightarrow 0$ on the value of $\gamma(0)$ is important in the subsequent analysis.

One of the main contributions of this paper is the analysis of our model for a broad class of system breach probability functions which are log-convex. Specifically, in Sec. \ref{sect:5.3}, we will define the class of functions $\gamma(s)$ for which our results hold, and it is broad and extends beyond constants and the constraint $\gamma(s) \leq 1$.

\subsection{Effort Complementarity}

The property of effort complementarity in (\ref{eq:2.6}) of $Z_{sv}$ is distinct from convexity. We digress to observe that such sign properties of second derivatives of independent variables have a long history in economics, for instance, in the form of the similar concept of ``cost complementarity". Cost complementarity in multi-product firms is the property that the marginal cost of producing product $j$ decreases with increased production of product $i$, where $i\neq j$. Baumol, Panzar, Willig's proof of the existence of ``natural monopoly" \cite{ref:Baumol, ref:Panzar1, ref:Braeutigam}, and Panzar and Willig's proof of the existence of subsidy-free prices \cite{ref:Panzar2, ref:Faulhaber} depend in part on cost complementarity.

\section{Attacker and Defender: Model, Actions, Reactions}
\label{sect3}

\subsection{Model of Attacker}
Let $y$ represent a generalized measure of the aggregate effort that the Attacker deploys. In the simplest representation, $y$ is the number of attempts, equal to the effort, that the Attacker makes to breach the Defender's system. Departures from this simple representation include breaching attempts of varying intensities. In the base representation where $y$ is the number of independent attacks, the total cost to the Attacker will be assumed to be $c \hspace{0.1em} y$, where $c$ is the Attacker's cost for unit effort. For a given system vulnerability $s$, we assume,
\begin{equation}\tag{3.1}
\label{eq:3.1}
Pr[\hspace{0.1em}\text{system is breached}\hspace{0.1em}] = T(s,y) = 1 - (1-s)^y
\end{equation}
The key assumption that we are making is that the Attacker's attempts are independent Bernoulli trials with probability of system breach in each trial given by $s$, and that the Attacker is successful if at least one of the attacks succeed in breaching the system.

We let $G$ denote the financial gain that the Attacker realizes if it succeeds. The Attacker's net expected gain for system vulnerability $s$ and effort $y$ in deployment is therefore,
\begin{equation}\tag{3.2}
\label{eq:3.2}
G \hspace{0.1em} T(s,y) - c \hspace{0.1em} y
\end{equation}

We are adopting features of the model in Danskin \cite{ref:Danskin}, who states (in Chapter 4, Sec.\hspace{0.2em}4), that the model was first posed during the Cold War at the RAND Corporation around 1951. In Danskin's model, there are several types of attack units indexed by $i$, the number of attack units employing weapons of type $i$ is denoted by $y_i$, the probability that an individual attack unit gets through the defenses $p_i$ is given by a specific exponential function that is monotonic, increasing with $y_i$, and the probability that the target survives the attack is $(1-p_i)^{y_i}$; our model is similar except there is a single type and the analog of $p_i$ is $s$, independent of $y_i$.

\subsection{Attacker's Optimization Problem}

The rational, risk-neutral Attacker will act to maximize its net expected gain, i.e., 
\begin{equation}\tag{3.3}
\label{eq:3.3}
\max_{y \geq 0} \ \Big[G \hspace{0.1em} T(s,y) - c \hspace{0.1em} y \Big]
\end{equation}
Let $y^*(s)$ denote the solution to the Attacker's optimization problem, and let $T^*(s) = T(s,y^*(s))$

\subsection{Model of Defender}

Our model assumes that the Defender has knowledge of the Attacker's decision process. We assume that the Defender is rational and risk-neutral in its decision-making, i.e., specifically, it is oblivious to risk, and hence its decisions are based on expected values only.

Let $L$ denote the financial loss to the Defender in the event of a system breach. Recall from Sec. \ref{sect2}, that $z$ denotes the Defender's aggregate effort at mitigating system vulnerabilities and $d$ is the unit cost of the effort, so that the net financial cost of mitigation is $d \hspace{0.1em} z$. Following the Attacker's actions and the Defender's mitigation efforts, the Defender's net expected financial loss is,
\begin{equation}\tag{3.4}
\label{eq:3.4}
L \hspace{0.1em} T^*(s) + d \hspace{0.1em} z
\end{equation}
which, upon substitution of the system breach probability function yields,
\begin{equation}\tag{3.5}
\label{eq:3.5}
L \hspace{0.1em} T^*(S(z,v)) + d \hspace{0.1em} z
\end{equation}

\subsection{Defender's Optimization Problem}
\label{sect:3.4}

The rational, risk-neutral Defender will implement the solution to the following problem,
\begin{equation}\tag{3.6}
\label{eq:3.6}
\min_{z \geq 0} \ \Big[L \hspace{0.1em} T^*(S(z,v)) + d \hspace{0.1em} z \Big]
\end{equation}
We find it convenient to conduct the analysis with the decision variable $s$ instead of $z$, which requires the replacement of the system breach probability function $S(z,v)$ by the effort function $Z(s,v)$, which has been introduced in Sec. \ref{sect:effort_function}. Since $z$ and $s$ are related by invertible functions, this transformation should not pose any fundamental problem. After the switch, the Defender's problem becomes,
\begin{equation}\tag{3.7}
\label{eq:3.7}
\min_{s \leq v} \ \Phi(s,v)
\end{equation}
where,
\begin{equation}\tag{3.8}
\label{eq:3.8}
\Phi(s,v) = L \hspace{0.1em} T^*(s) + d \hspace{0.1em} Z(s,v)
\end{equation}
We denote the solution to (\ref{eq:3.7}) by $s^*(v)$, also $z^*(v)=Z(s^*(v),v)$, which gives the Defender's optimum mitigation effort, and $\Phi^*(v)=\Phi(s^*(v),v)$.

\section{Attacker's Problem: Solution and Discussion}
\label{sect4}

\subsection{Solution}
\label{sect:4.1}

The first order condition for optimality in the problem in (\ref{eq:3.3}), which is obtained by setting to zero the derivative of the objective function, yields,
\begin{equation}\tag{4.1}
\label{eq:4.1}
\{ -\log(1-s)\}(1-s)^y = \frac{c}{G}
\end{equation}
The left-hand side is monotonic, decreasing in $y$ and approaches 0 as $y \rightarrow \infty$. Hence, if its value for $y=0$ is greater than $c/G$, then a unique positive solution $y^*(s)$ exists, and otherwise the solution to the Attacker's problem in (\ref{eq:3.3}) is $y^*(s)=0$.

Define $s_P$ to be the value of $s$ such that the left-hand side of (\ref{eq:4.1}) at $y=0$ is equal to $c/G$, i.e,
\begin{equation}\tag{4.2}
\label{eq:4.2}
s_P = 1 - e^{-c/G}
\end{equation}
We summarize here:

 \begin{proposition} \label{prop:sp}
If $s > s_P$, then a unique positive solution to $y^*(s)$ to the Attacker's Optimization Problem in (\ref{eq:3.3}) exists, and satisfies,
 \end{proposition}
\begin{equation}\tag{4.3, i} 
\label{eq:4.3.1}
(1-s)^{y^*(s)} = \frac{1}{(G/c)\{-\log(1-s)\}}
\end{equation}
i.e.,
\begin{equation}\tag{4.3, ii}
\label{eq:4.3.2}
y^*(s) = -\frac{\log \left[ \log \big\{(1-s)^{-G/c}\big\} \right]}{\log(1-s)}
\end{equation}
If $s \leq s_P$, then $y^*(s)=0$.

Also, the probability of a system breach that is a consequence of the Attacker's optimal effort,
\begin{align}
\label{eq:4.4}
T^*(s) &= 1 + \frac{1}{(G/c)\{\log(1-s)\}} \qquad \text{for } s > s_P \tag{4.4, i}\\
&= 0 \hspace{11.2em} \text{for } s \leq s_P \tag{4.4, ii}
\end{align}
Observe that,
\begin{equation}\tag{4.5}
\label{eq:4.5}
\frac{d \hspace{0.1em} T^*(s)}{ds} > 0 \qquad \text{for } s > s_P
\end{equation}

In the Attacker's problem, the system breach probability $s$ is given and assumed to satisfy the constraint $s=S(z,v) \leq v$, the initial vulnerability. Note that if $v \leq s_P$ then $s \leq s_P$. To avoid trivialities, we assume that $v > s_P$.

\vspace{10px}

\subsection{Discussion, the Price of Deterrence}
\label{sect:4.2}

We examine the behavior of $y^*(s)$, the Attacker's optimal effort as a function of $s$, the system vulnerability, for $s > s_P$. 

\begin{figure}[ht]%
    \figuretag{4.1}
    \captionsetup{justification=centering}
    \centering
    \begin{subfigure}{6.5cm}
    \includegraphics[width=\linewidth]{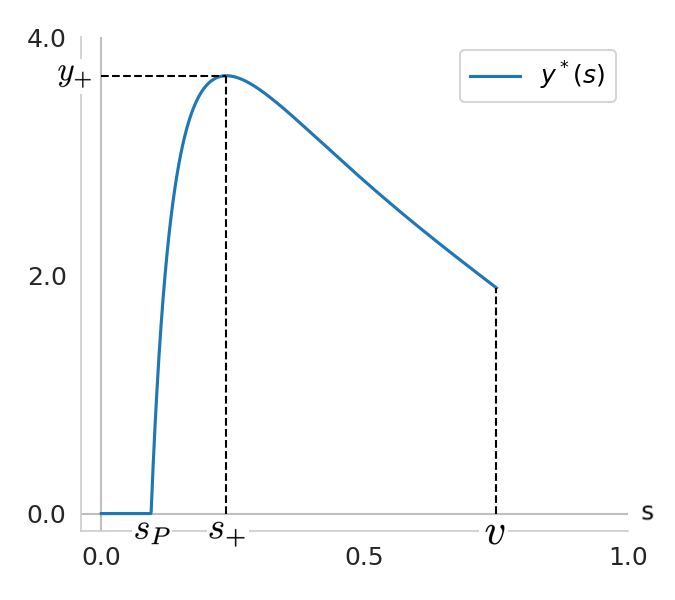}
    \caption{$y^*(s)$ vs. $s$}\label{fig:4.1.1}
    \end{subfigure}
    \qquad
    \begin{subfigure}{6.5cm}
    \includegraphics[width=6.5cm]{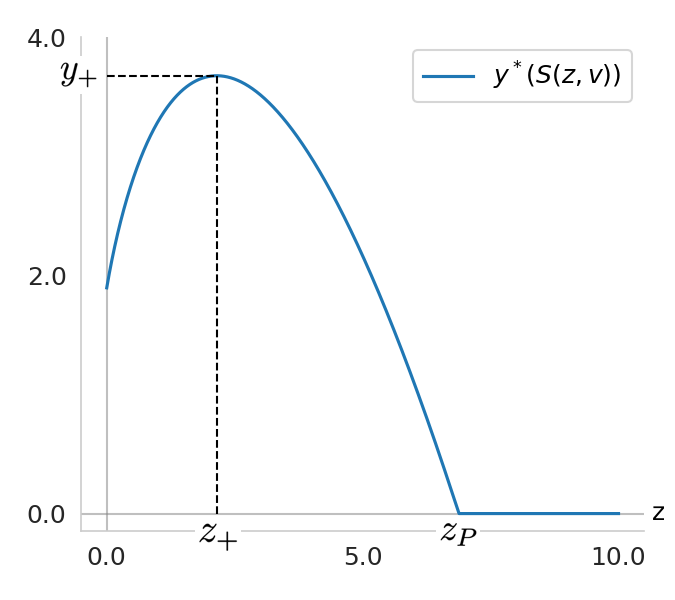}
    \caption{$y^*(S(z,v))$ vs. $z$}\label{fig:4.1.2}
    \end{subfigure}
    \caption{Visualizing rational Attacker investments and the Price of Deterrence;\\GL Class I, $v=0.75$, $L=G=10$, $\alpha=1$, $\beta=1$, $c=d=1$; $\longrightarrow$ $y_+ \approx 3.68$, $s_+ \approx 0.24$, $z_+ \approx 6.88$} %
    \label{fig:y_s}%
\end{figure}

Differentiating the expression in (\ref{eq:4.3.2}) with respect to $s$ yields an expression for $\frac{dy^*(s)}{ds}$ from which it may be inferred (the detailed proof is omitted), for $s > s_P$, that as $s$ increases, $y^*(s)$ is initially monotonically increasing, reaches a stationary point, which we denote by $s_+$, and is thereafter monotonic, decreasing, as depicted in Fig. \ref{fig:y_s}. We let the maximum value of $y^*(s)$ be $y_+$, i.e., $y_+ = y(s_+)$. The following result explicitly identifies $s_+$ and $y_+$.

 \begin{proposition} \label{prop:expDeccase}
 The Attacker's maximum effort over all system vulnerabilities, $y_+$, is $1/e$, i.e., about 37\% of the effort justified by the potential gain from system breach. This follows from,
 \end{proposition}
\begin{equation}\tag{4.6}
 s_+ = 1-e^{-ec/G}, \qquad y_+ =\frac{G/c}{e}, \qquad \text{so that } \ (1-s_+)^{y_+} = \frac{1}{e}
\end{equation}
\vspace{-3px}

This result resembles Gordon-Loeb's $1/e$ result, which follows from the assumption of log convexity of $S(z,v)$ with respect to $z$. Here $T(s,y)$ in (\ref{eq:3.1}) is log-concave with respect to $y$, and yields the similar $1/e$ result for the Attacker's optimization problem in (\ref{eq:3.2}). 
The corresponding behavior of $y^*(Z(s,v),v)$ as a function of the Defender's effort $z$ is shown in Fig. \ref{fig:4.1.2}. In this figure, 
\begin{equation}\tag{4.7}
\label{eq:4.7}
z_P = Z(s_P, v), \quad \text{i.e.}, \quad S(z_P,v) = s_P,
\end{equation}
and $z_+$ is defined similarly from $s_+$. When $G/c$ is large, as may be considered typical, then from (\ref{eq:4.2}), $(s_P)(G/c) \approx 1$. Now, $s_P$ is the probability of system breach from the Attacker's unit effort, and $G/c$ is the maximum effort that a rational Attacker will deploy. Hence, in this approximation regime, when $s \leq s_P$, the expected number of successful system breaches falls below one even when the Attacker mounts the maximum effort that is justified by the potential gain from a system breach. Consequently, in this case the rational, risk-neutral Attacker will not mount \textit{any} attack.

Turning to the implications for the Defender's strategy, we focus on $z_P$, see Fig. \ref{fig:4.1.2}, the Defender's effort that yields the system breach probability of $s_P$. We call $z_P$ the ``Price of Deterrence" (strictly it is $d$\hspace{0.1em}$z_P$). That is, if $z > z_P$, then the system breach probability falls below the threshold that justifies the Attacker to mount any attack.

It is important to keep in mind that the Price of Deterrence depends on the initial vulnerability $v$. This is evident in (\ref{eq:4.7}): $G/c$ determines $s_P$, but $z_P$ depends on $v$ as well.

\subsection{Arms Race Analogy}

We draw attention to the Price of Deterrence, $z_P$ in Fig. \ref{fig:4.1.2}: as the Defender's investment $z$ increases, the Attacker's optimum effort, $y^*$, initially increases, i.e., the protagonists are locked in a classic arms race. However, if the Defender's investment exceeds $z_+$, the Attacker's economic interests no longer justify matching the escalation, and its investments decline. If the Defender's investment reaches the Price of Deterrence, then the Attacker throws in the towel, and cuts off all further investments. Note that the entire cycle consisting of the Attacker's expanding investments, followed by the partial withdrawal and finally total withdrawal while facing a Defender with superior economic resources is symptomatic of the point of view that economics is a substantial strategic factor in cybersecurity.

It is tempting to conjecture that the behavior in Fig. \ref{fig:4.1.2} mirrors the US -- Soviet Union arms race during the Cold War until the economic collapse and dissolution of the Soviet Union (``Part of the logic proceeding with SDI was that, eventually, the arms race would cripple the Soviet economy. This is in fact what was happening." \cite{ref:Swift}). See also \cite{ref:Goodman}.

\section{Defender's Problem: Analysis and Solution}
\label{sect5}

\subsection{Objective and Gradient Functions}

Combining (\ref{eq:3.8}) and (\ref{eq:4.4}) yields the following expression for the Defender's objective function for $s > s_P$,
\begin{equation}\tag{5.1} 
\label{eq:5.1}
\Phi(s,v) = L \left[ 1 + \frac{1}{(G/c)\{\log(1-s) \}}\right] + d\hspace{0.1em} Z(s,v)
\end{equation}
In deriving the derivative with respect to $s$, the system breach probability, we take note of the factored form of $Z_s(s,v) = -\frac{f(v)}{g(s)}$ in (\ref{eq:2.7}), to obtain,
\begin{equation}\tag{5.2, i} 
\label{eq:5.2.1}
\left[ \frac{(1-s)\log^2(1-s)}{d\hspace{0.1em}f(v)}\right] \frac{\partial}{\partial s} \Phi(s,v) = \frac{L/d}{G/c} \frac{1}{f(v)} - D(s)
\end{equation}
where,
\begin{equation}\tag{5.2, ii} 
\label{eq:5.2.2}
D(s) = \frac{(1-s)\log^2(1-s)}{g(s)}
\end{equation}
Note that the bracketed term on the left-hand side of (\ref{eq:5.2.1}) is positive for $0 < s < 1$, and the separation of $v$ and $s$ in the terms on the right-hand side. Since the system parameter $\frac{L/d}{G/c}$ has an important role in the analysis, we let,
\begin{equation}\tag{5.3} 
\label{eq:5.3}
R = \frac{L/d}{G/c}
\end{equation}
and refer to $R$ as the Effective Loss to Gain Ratio, where it is understood that the (potential) loss and gain are of the Defender and Attacker, respectively. Now,
\begin{equation}\tag{5.4}
\begin{aligned}
\label{eq:5.4}
\Phi_s(s,v) &= 0 \qquad \text{if } \quad D(s) = R \frac{1}{f(v)} \\[3px]
          &> 0 \qquad \text{if } \quad D(s) < R \frac{1}{f(v)} \\[3px]
          &< 0 \qquad \text{if } \quad D(s) > R \frac{1}{f(v)} \\[3px]
\end{aligned}
\end{equation}
Though $D(s)$ is only a component of the gradient with respect to $s$ of the Defender's two-variable objective function, we will take the liberty of referring to it as the ``gradient function".

\subsection{Example}
\label{sect:5.2}

We introduce the ``universal" function $\xi(s)$ here to illustrate a gradient function and also because it will be useful later in the analysis:
\begin{equation}\tag{5.5} 
\label{eq:5.5}
\xi(s) = \frac{(1-s)\log^2(1-s)}{s}
\end{equation}
For GL Class II functions, for which $g(s) = \alpha s$, see (\ref{eq:2.13}), the gradient function $D(s) = \frac{\xi(s)}{\alpha}$. In the interval $[0,1]$, the corner points of $\xi(s)$ are $\xi(0) = \xi(1) = 0$. The function $\xi(s)$ has the following canonical shape: a unique maximum $\hat{\xi} = \max_{0 < s < 1} \xi(s)$, monotonic, increasing behavior in $[\hspace{0.1em}0, \hat{s}_\xi)$, and monotonic, decreasing behavior in $(\hat{s}_\xi$,1\hspace{0.1em}], where we have denoted the location of the maximum by $\hat{s}_\xi$, i.e., $\xi(\hat{s}_\xi)=\hat{\xi}$. We shall say that such functions have an ``inverted-U shape"\footnote{Use of the term ``inverted-U" function or relationship has precedents, for example, see \cite{ref:Aghion}.}. In the example of $\xi(s)$ in (\ref{eq:5.5}), $\hat{\xi}  \approx 0.64$ and $\hat{s}_\xi \approx 0.8$. Note that such functions are not necessarily concave, even though some properties are shared.

\subsection{Shape of the Gradient Function}
\label{sect:5.3}

Here we obtain the shape of the gradient function $D(s)$. A key determinant is the non-negative function $\gamma(s)$, which was introduced in (\ref{eq:2.16}) to parameterize the log convexity of the system breach probability function $S(.,.)$. Taking the derivative of $D(s)$ in (\ref{eq:5.2.2}) with respect to $s$, and making use of (\ref{eq:2.17}) gives,
\begin{equation}\tag{5.6} 
\label{eq:5.6}
\left[ \frac{\hspace{0.1em} g(s)}{\log^2(1-s)} \right]\frac{d}{d s} D(s) = H(s; \gamma)
\end{equation}
where the bracketed term on the left-hand side is positive for $s \in (0,1)$, and ($H$ for \textit{Hessian}),
\begin{equation}\tag{5.7} 
\label{eq:5.7}
H(s; \gamma) = - \left\{ \frac{2}{\log(1-s)} + \frac{1}{s}\right\} - \gamma(s) \left(\frac{1}{s}-1\right)
\end{equation}
We proceed below with separate analyses for $0 \leq \gamma(s) \leq 1$ and $1 < \gamma(s), \ s \in [0,1]$. Let,
\begin{equation*}
\Gamma_1 = \{ \gamma(s): 0 \leq \gamma(s) \leq 1, \ \text{and} \ \ 0 \leq \gamma_s(s), \ \forall s \in [\hspace{0.1em}0,1] \}
\end{equation*}
Observe that $\gamma(s)$ corresponding to GL Class I functions with $\gamma \leq 1$, and GL Class II functions are subsumed in the set $\Gamma_1$.
\begin{proposition} \label{prop:5.1}
If $\gamma(s) \in \Gamma_1$ then the Hessian $H(s; \gamma)$ monotonically decreases, i.e., $H_s(s;\gamma) < 0, \ s \in (0,1)$, from non-negative to negative with increasing $s, \ s \in [0,1]$. Hence, $\exists$ unique $\hat{s} \in (0,1)$ such that $H(\hat{s};\gamma)=0$, where $D_s(\hat{s})=0$ and $D_s(s) > (<) \ 0$ for $s < (>) \ \hat{s}$. In the special case of $\gamma(s) \equiv 1$, $\hat{s}=0$.
\end{proposition}

\vspace{5px}

The proof is in Appendix \nameref{app:A1}. From (\ref{eq:5.6}), $D(\hat{s})= \max_{0 \leq s \leq 1} D(s)$, and we let $\hat{D}=D(\hat{s})$. The proposition establishes that for $\gamma(s) \in \Gamma_1$, the gradient function $D(s)$ has the inverted-U shape with the peak value of $\hat{D}$ at $\hat{s}$. In the special case of $\gamma(s) \equiv 1$, the peak is located at $\hat{s}=0$, and the shape of the entire gradient function $D(s), s \geq 0$, coincides with the segment of the general inverted-U shape to the right of the peak, i.e., $s \geq \hat{s}$.

We can prove that in general $D(s)$ is not concave. This fact gives weight to the gradient function's inverted-U shape since this property will prove to be adequate for deducing key properties of stationary points of the Defender's objective function, which is considered in Sec. \ref{sect:5.5}.

Next, we investigate the behavior of the gradient function $D(s)$ when the characteristics of the function $\gamma(s)$ are complementary to that of Prop. \ref{prop:5.1}. Let,
\begin{equation*}
\Gamma_2 = \{ \gamma(s): 1 < \gamma(s), \ \forall s \in [\hspace{0.1em}0,1] \}
\end{equation*}
Observe that $\gamma(s)$ corresponding to GL Class I functions with $\gamma > 1$ are subsumed in $\Gamma_2$.
\begin{proposition} \label{prop:5.2}
If $\gamma(s) \in \Gamma_2$, then $H(s; \gamma) < 0$ and $D_s(s) < 0, \ \forall s \in (0,1)$. Also, $D(s) \rightarrow \infty$ as $s \rightarrow 0$.
\end{proposition}

\vspace{5px}

The proof is in Appendix \nameref{app:A1}. For $\gamma(s) \in \Gamma_2$, $D(s) \rightarrow \infty$ as $s \rightarrow 0$ which represents a significant qualitative difference from $\gamma(s) \in \Gamma_1$; $D(s)$ is monotonic, strictly decreasing for $s > 0$. We let $\hat{s}=0$.

We now build on the results of Prop. \ref{prop:5.1} and \ref{prop:5.2} to obtain the shape of the gradient function for a more general class of functions $\gamma(s)$,
\begin{equation}\tag{5.8}
\label{eq:5.8}
\Gamma = \{ \gamma(s): \gamma(s) \geq 0, \ \text{and} \ \ \gamma(s) \leq 1 \Rightarrow \gamma_s(s) \geq 0, \ s \in [\hspace{0.1em}0, 1] \}
\end{equation}

An example of $\gamma(s) \in \Gamma$ is,
\begin{equation}\tag{5.9}
\label{eq:5.9}
\gamma(s) = -\frac{1}{2}s^2 + \frac{3}{4}s + \frac{7}{8}
\end{equation}

Note that $\gamma(0) < 1, \gamma(s)$ is monotonic, increasing while $\gamma(s) \leq 1$, the $\gamma=1$ level is crossed at $s_c=(3-\sqrt{5})/4 \approx 0.19$, and for $s_c < s \leq 1$, $\gamma(s) > 1$. Hence, in this example, $\gamma(s)$ behaves initially as an element of $\Gamma_1$ and as an element of $\Gamma_2$ after the cross-over at $s_c$.

We will now analyze the class of functions in $\Gamma$ allowing for the possibility of a single cross-over of behavior from $\Gamma_1$ to $\Gamma_2$ to avoid getting immersed in details. Note that if for any $s'$, $\gamma(s') > 1$, then no crossing below the $\gamma=1$ level is possible for $s > s'$.

Now suppose $\gamma(0) \leq 1$. If $\gamma(s) \leq 1, \ \forall s \in [\hspace{0.1em}0,1]$, then Prop. \ref{prop:5.1} will apply. Next, suppose the following alternative: $\gamma(0) < 1$ and $\exists \ s_c, s_c \in (0,1]$ such that $\gamma(s) < 1$ for $s < s_c$ and $\gamma(s) > 1$ for $s > s_c$, i.e., there is a cross-over at $s_c$. In this case, $\gamma(s) > 1, \ \forall s \in [\hspace{0.1em}s_c, 1]$. Continuing with this case,
\begin{enumerate}[label=(\roman*),itemsep=-3px]
    \item Since $\gamma(s) > 1, \ s\in (s_c,1]$, from Prop. \ref{prop:5.2}, $H(s; \gamma) < 0$. In particular, $H(s_c; \gamma) < 0$.
    \item For $s < s_c$, since $\gamma(s) \leq 1$, and $\gamma_s(s) \geq 0$, from Prop. \ref{prop:5.1}, $H_s(s;\gamma) < 0, \ s \in [\hspace{0.1em}0, s_c]$.
\end{enumerate}
Combining (i) and (ii), since $H(0;\gamma)>0$ and $H(s_c;\gamma) < 0$, it follows that $\exists \ s_B, \ s_B < s_c$, such that $H(s_B, \gamma)$ = 0. Hence, from (\ref{eq:5.6}),
\begin{equation}\tag{5.10}
\label{eq:5.10}
\frac{d\hspace{0.1em}D(s_B)}{ds} = 0
\end{equation}

Since $D_s(s) > (<) \ 0$ for $s < (>) \ s_B$, $D(s)$ is inverted-U shaped with its maximum at $s_B$.

\begin{proposition}
\label{prop:5.3}
Assume that $\gamma(s) \in \Gamma$. If $\gamma(0) > 1$, then Prop. \ref{prop:5.2} applies. If $\gamma(0) \leq 1$, then $D(s)$ is inverted-U shaped with a unique maximum at $\hat{s} = s_B$.
\end{proposition}

An interpretation of the proposition is that if $\gamma(0) > 1$, then the shape of the gradient function is what is expected from Prop. \ref{prop:5.2} for $\gamma(s) \in \Gamma_2$, and if $\gamma(0) \leq 1$, then the gradient function is inverted-U shaped, as stated in Prop. \ref{prop:5.1} for $\gamma(s) \in \Gamma_1$.

\subsubsection{Examples} 
\begin{figure}[ht]%
    \figuretag{5.1}
    \captionsetup{justification=centering}
    \centering
    \begin{subfigure}{6cm}
    \includegraphics[width=\linewidth]{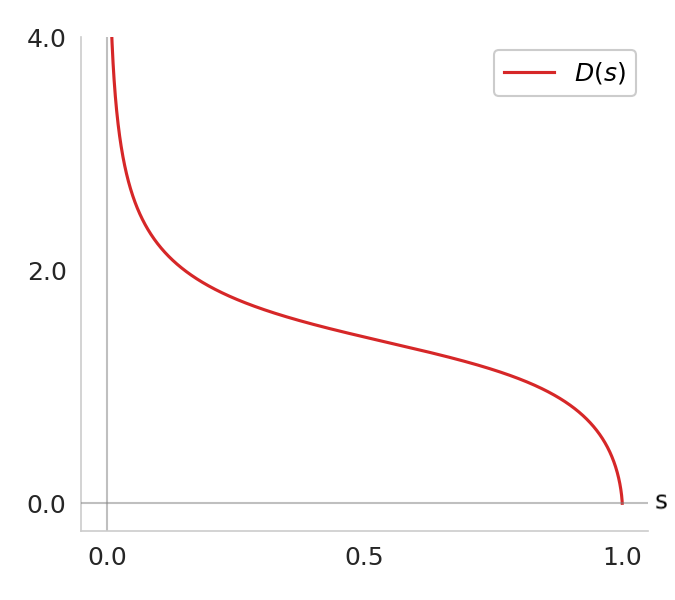}
    \caption{GL Class I ($\gamma = 1.25$)}\label{fig:5.1.1}
    \end{subfigure}
    \qquad
    \begin{subfigure}{6cm}
    \includegraphics[width=6cm]{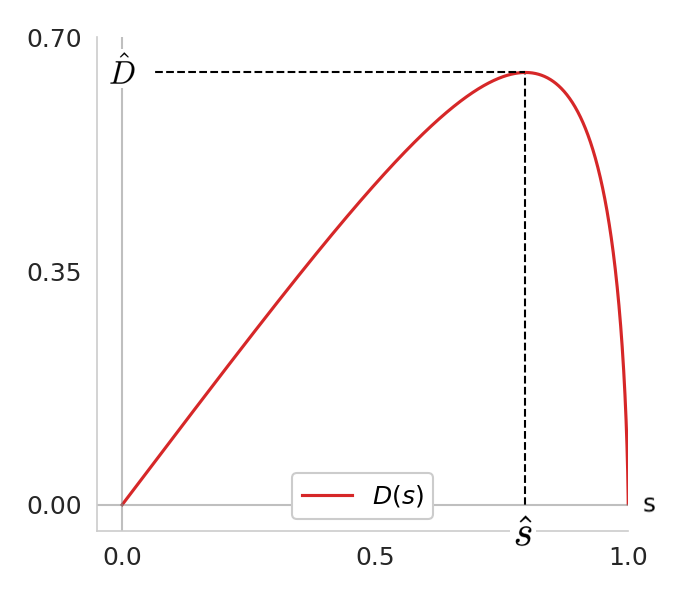}
    \caption{GL Class II ($\gamma = 0$)}\label{fig:5.1.2}
    \end{subfigure}
    \caption{Visualizing $D(s)$ for GL Class I and II functions. The GL Class I function with $\gamma < 1$, i.e., $\beta > 1$, is given in Fig. \ref{fig:stationary_points}.}%
    \label{fig:ds}%
\end{figure}

In Fig \ref{fig:ds}, for the GL Class I function ($\gamma = 1/\beta$) with $\alpha=1.0$, $\beta=0.8$ ($\gamma = 1.25$), $\hat{s}=0$, $\hat{D}=+\infty$, and for the GL Class II function ($\gamma = 0$) with $\alpha=1.0$, $\hat{s} \approx 0.8$, $\hat{D} \approx 0.64$.

\subsection{Stationary Points of the Defender's Objective Function}

The stationary points, i.e., local maxima and minima, of the Defender's objective function $\Phi(s,v)$ for the variable $s$ and fixed $v$, the initial vulnerability, are solutions of $\Phi_s(s,v)=0$. Therefore, following (\ref{eq:5.2.1}), the stationary points are solutions of,
\begin{equation}\tag{5.11} 
\label{eq:5.11}
D(s) = R \frac{1}{f(v)} \\[5px]
\end{equation}
If $\gamma(s) \in \Gamma_2$, then from Prop. \ref{prop:5.2} it is known that $D(s) \rightarrow \infty$ as $s \rightarrow 0$ and also that $D(s)$ is monotone, strictly decreasing with increasing $s$. Hence, a unique solution $s_1(v), \ 0 < s_1(v) < 1$, exists. We next focus on $\gamma(s) \in \Gamma_1$ for which, from Prop. \ref{prop:5.1}, $D(s)$ is inverted-U shaped. Since $f(v)$ increases with $v$, for any solution of (\ref{eq:5.11}) to exist, the initial vulnerability $v$ needs to be sufficiently high. In fact, any stationary point exists if and only if,
\begin{equation}\tag{5.12} 
\label{eq:5.12}
\hat{v} \leq v
\end{equation}
where $\hat{v}$ is defined by the following relation,
\begin{equation}\tag{5.13} 
\label{eq:5.13}
\hat{D} = R\frac{1}{f(\hat{v})}, \ \ \ \text{i.e.,} \ \ \ \hat{v} = f^{-1} \left( \frac{R}{\hat{D}} \right)
\end{equation}
and $\hat{D}$ is the maximum value of $D(s)$. 
\subsubsection*{Examples}
\begin{enumerate}[label=(\roman*),itemsep=-5px]
    \item For GL Class I functions, $f(v)=v^{1/\beta}$, see (\ref{eq:2.10}), so that, for $\gamma \leq 1$, i.e., $\beta \geq 1$,
    \begin{equation}\tag{5.14} 
    \label{eq:5.14}
    \hat{v} = \left(\frac{R}{\hat{D}} \right)^\beta
    \end{equation}
    \item For GL Class II functions, recall from Sec. \ref{sect:5.2} that $\hat{D} = \frac{\hat{\xi}}{\alpha}$, and $f(v) = -\frac{1}{\log(v)}$, hence,
    \begin{equation}\tag{5.15} 
    \label{eq:5.15}
    \hat{v} = \exp \left( -\frac{\hat{\xi}}{\alpha R} \right)
    \end{equation}    
\end{enumerate}

Observe from the above expressions that for both Class I with $\gamma \leq 1$ and Class II functions, $\hat{v}$ is independent of initial vulnerability $v$, and importantly, increases with the effective loss to gain ratio, $R$.

If $\gamma(s) \in \Gamma$ and $\gamma(0)<1$, then there exist two stationary points of $\Phi(s,v)$, denoted $s_2(v)$ and $s_1(v)$, where,
\begin{equation}\tag{5.16}
\label{eq:5.16}
s_2(v) < \hat{s} < s_1(v) \quad \text{if } \hat{v} < v,
\end{equation}
and, in the special case of $v=\hat{v}$, \ $s_2(v) = \hat{s} = s_1(v)$. It is easy to see that $s_1(v)$ ($s_2(v)$) increases (decreases) with $v$. 

If $\gamma(s) \in \Gamma$ and $\gamma(0) > 1$, then there is a single stationary point of $\Phi(s,v)$, denoted by $s_1(v)$ which increases with $v$. Hereafter it is convenient to consider only $\gamma(s) \in \Gamma$ and $\gamma(0) \leq 1$, and accommodate the case of $\gamma(s) \in \Gamma$ and $\gamma(0) > 1$ by noting that in the latter case $\hat{s}=0$, and consequently $s_2(v)$ does not exist in $[\hspace{0.1em}0,1]$.

Using (\ref{eq:5.2.1}) to obtain the sign of $\Phi_s(s,v)$ in the intervals defined by the stationary points, we obtain,

\begin{proposition} \label{prop:5.4}
Assume $\gamma(s) \in \Gamma$ and $v$ is constant. Then,
\end{proposition}
\vspace{-10px}
\begin{equation}\tag{5.17}
\begin{aligned}
\label{eq:5.17}
\Phi_s(s,v) &> 0 \quad \text{for } s \in (0, s_2(v)) \\
&< 0 \quad \text{for } s \in (s_2(v), s_1(v)) \\
&> 0 \quad \text{for } s \in (s_1(v), 1)
\end{aligned}
\end{equation}

That is, $s_2(v)$ is a local maximum and $s_1(v)$ is a local minimum of $\Phi(s,v)$ as $s$ is varied in $(0,1)$ with $v$ fixed.

\begin{figure}[ht]%
    \figuretag{5.2}
    \centering
    \captionsetup{justification=centering}
    \begin{subfigure}{6cm}
    \includegraphics[width=\linewidth]{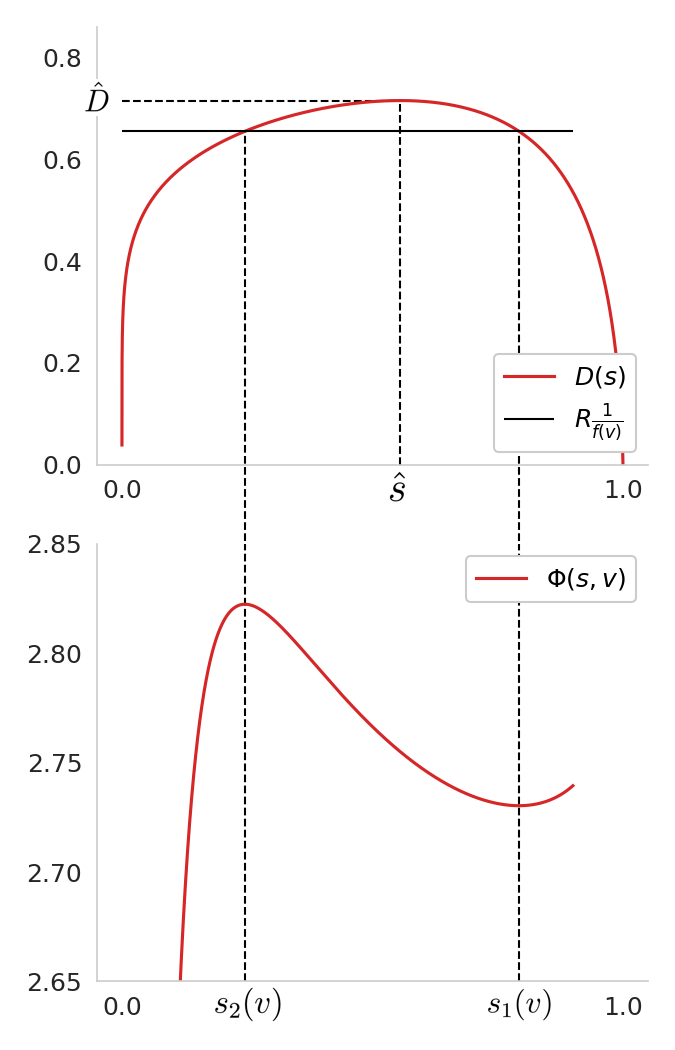}
    \caption{$v = 0.90$}\label{fig:5.2.1}
    \end{subfigure}
    \qquad
    \begin{subfigure}{6cm}
    \includegraphics[width=6cm]{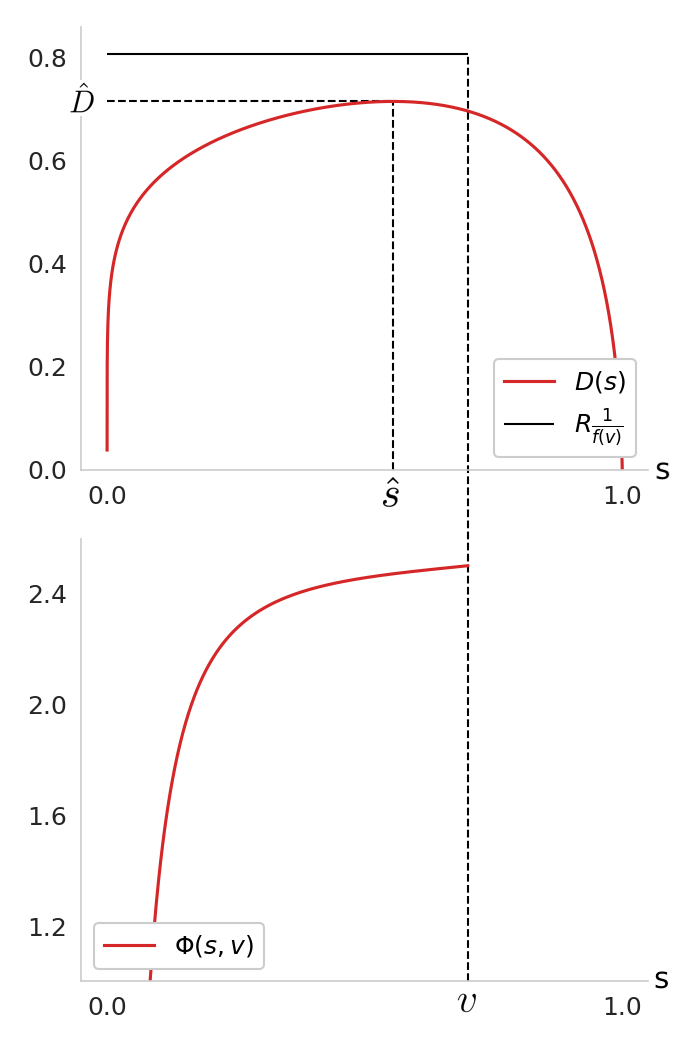}
    \caption{$v = 0.70$}\label{fig:5.2.2}
    \end{subfigure}
    \caption{Visualizing criteria for the existence of stationary points in the Defender's objective function; GL Class I, $R=0.6,\ \alpha=1,\ \beta=1.2$}%
    \label{fig:stationary_points}%
\end{figure}

The above result illuminates the main reason for complexity of the analysis in this paper, i.e., the combination of intervals of local convexity and concavity in one variable ($s$), with initial vulnerability $v$ adding a separate dimension of complexity.

\subsection{Solution to the Defender's Problem}
\label{sect:5.5}

We compose the solution to the problem stated in (\ref{eq:3.7}). The main task is to deduce the implications of the constraint $s \leq v$ in the Defender's problem in (\ref{eq:3.7}), in combination with the constraint-free result in Prop. \ref{prop:5.4}.

If $v < \hat{v}$ then, from (\ref{eq:5.2.1}), $\Phi_s(s,v) > 0, \ \forall s$, and,
\begin{equation}\tag{5.18} 
\label{eq:5.18}
\Phi^*(v) = \Phi(s_P,v) = d \hspace{0.1em} z_P
\end{equation}  
Recall from the discussion in Sec. \ref{sect:4.2} that $d \hspace{0.1em} z_P$ is the Price of Deterrence. If $v > \hat{v}$, then from Prop. \ref{prop:5.4}, we need to consider two disjoint cases, $v < s_1(v)$,and $s_1(v) < v$. From Prop. \ref{prop:5.4},
\begin{equation}\tag{5.19} 
\label{eq:5.19}
\text{If} \ \ \hat{v} < v < s_1(v) \ \ \text{then} \ \ \Phi^*(v)=\min \left[ \Phi(s_P,v), \Phi(v,v) \right]
\end{equation}  
so that $s^*=s_P$ or $v$. Since $\Phi(v,v)$ is obtained from $s=v$, i.e., $z=0$,
\begin{equation}\tag{5.20} 
\label{eq:5.20}
\Phi(v,v) = L\hspace{0.1em}T^*(v) = L \left[ 1 + \frac{1}{(G/c)\log(1-v)}\right],
\end{equation} 
a positive quantity since we have assumed (see Sec. \ref{sect:4.1}) that $v > s_P$. Note that $\Phi(v,v)$ is the Defender's net expected loss if it makes no effort to reduce vulnerability from its initial value, and $\Phi(s_P,v)$ is the corresponding quantity if it makes the effort necessary to reduce vulnerability to the level where the Attacker is deterred from making any effort. It is not obvious a priori which is the superior solution for the Defender. Hence we have (\ref{eq:5.19}).

In contrast,
\begin{equation}\tag{5.21} 
\label{eq:5.21}
\text{If} \ \ s_1(v) < v \ \ \text{then} \ \ \Phi^*(v) = \min \left[ \Phi(s_P,v), \Phi(s_1(v),v) \right]
\end{equation} 
so that $s^*=s_P$ or $s_1(v)$.

We define three ``Decision Intervals" (DI) in the range of values of $v$:
\begin{enumerate}[leftmargin=0.53in,label=DI\hspace{0.5em}\arabic*,itemsep=-4px,ref=5.22]
    \item \hspace{-4px}. $v < \hat{v}$
    \item \label{enum:5.22} \hspace{-4px}. $\hat{v} < v < s_1(v)$ \hspace*{\fill}(5.22)
    \item \hspace{-4px}. $s_1(v) < v$
\end{enumerate}

DI 2 and 3 may be empty, e.g., DI 3 is empty if $v \leq s_1(v), \ \forall v$. Also, as we shall see, the DI may be interleaved. To describe the Defender's investment decision, we employ the following mnemonics:
\begin{enumerate}[itemsep=-3px]
    \item ``all in" to describe $\Phi^*(v)=\Phi(s_P,v)$
    \item ``none" if $\Phi^*(v)=\Phi(v,v)$
    \item ``some" if $\Phi^*(v)=\Phi(s_1(v),v)$
\end{enumerate}
Now, by combining Prop. \ref{prop:5.4} and the constraint $s \leq v$, we obtain the following result for the Defender's problem in (\ref{eq:3.7}),

\vspace{5px}

\begin{proposition} \label{prop:5.5}
The Defender's optimal investment decision depends on the DI in which $v$ exists, as follows: DI 1: ``all in; DI 2: ``all in" or ``none"; DI 3: ``all in" or ``some".
\end{proposition}

Note the role of the parameter $R$, the effective loss to gain ratio, in the Defender's decision. For instance, since $\hat{v}$ is monotonic, increasing with $R$, if $R$ (and therefore $\hat{v}$) is sufficiently large then for all cases of practical interest the initial vulnerability $v < \hat{v}$, in which case only Interval 1 is of interest, and the Defender's decision is ``all in".

The above observation on the role of $R$ also illuminates the potential of spreading false estimates of loss ($L$) and gain ($G$) to nudge the adversary to making damaging decisions. This is a rich line of investigation that, however, is outside the scope of this paper.

\section{Defender's Optimal Investment Policy}
\label{sect6}

The preceding section has shown how to obtain the Defender's optimal investment for any given value of the initial vulnerability, $v$. Here, we investigate patterns in optimal investment decisions for the entire range of values of $v$. Naturally this also involves investigating dependence on parameters, such as $R$, the effective loss to gain ratio. From the reader's perspective, an important and illuminating component of this section is the illustrations of results and properties related to the analysis. The analysis and illustrations in this section are for Gordon-Loeb's Class I and II system breach probability functions. Lastly, this section compares and contrasts the optimum investment policies that are obtained from the model of Gordon-Loeb and our model.

\subsection{Fixed Point Equations}

Section \ref{sect:5.5}, and specifically (\ref{enum:5.22}), has shown that the transitions between Decision Intervals (DI) 2 and 3 occur when $\{s_1(v)-v\}$ changes sign, where $v$ is the initial vulnerability and $s_1(v)$ is the unique local minimum of the Defender's objective function $\Phi(s,v)$ when $s$, the system breach probability, is varied with $v$ held constant. Hence, it is natural to investigate the condition in which equality holds, and this leads to the Fixed Point Equation (FPE) below. Since $s_1(v)$, when it exists, is obtained as the solution of the equation $D(s_1(v))=\frac{R}{f(v)}$, $s_1(v) \geq \hat{s}$, the transition points in $v$, i.e., solutions of $s_1(v)=v$, are solutions of the following FPE,
\begin{equation}\tag{6.1} 
\label{eq:6.1}
D(x)=\frac{R}{f(x)}, \ \ \hat{s} < x < 1
\end{equation} 
Note that any solution of (\ref{eq:6.1}) will satisfy $x > \hat{v}$. Since we will not encounter more than two Fixed Points, we denote the solutions of (\ref{eq:6.1}) by $v^L$ and $v^H$. If two solutions exist, we assume that $\hat{v} < v^L < v^H$, and if there is only one, then $\hat{v} < v^H$.

\subsection{Gordon-Loeb Class I System Breach Probability Functions}

In this case, the FPE is,
\begin{equation}\tag{6.2, i} 
\label{eq:6.2.1}
\frac{(1-x)\log^2(1-x)}{\alpha \beta x^{(\beta+1)/\beta}} = \frac{R}{x^{1/\beta}}, \ \ \hat{s} < x < 1,
\end{equation}
or equivalently,
\begin{equation}\tag{6.2, ii} 
\label{eq:6.2.2}
\xi(x) = R\alpha \beta, \ \ \hat{s} < x < 1
\end{equation}
where $\xi(.)$ is the ``universal" function defined in (\ref{eq:5.5}). 

\begin{proposition}
\label{prop:6.1}
For GL Class I functions,
\begin{enumerate}[label=(\roman*),itemsep=-3px]
    \item If $\hat{\xi} < R \alpha \beta$, then no Fixed Point exists, and $s_1(v) < v, \ \forall v \in (\hat{v},1)$.
    \item If $\xi(\hat{s}) < R \alpha \beta < \hat{\xi}$, then two Fixed Points, $v^H$ and $v^L$ exist, and,
    \begin{align*}
        s_1(v) < v &, \ \forall v \in (\hat{v}, v^L) \\
        v < s_1(v) &, \ \forall v \in (v^L, v^H) \\
        s_1(v) < v &, \ \forall v \in (v^H, 1)
    \end{align*}
    
    \item If $R \alpha \beta < \xi(\hat{s})$, then one Fixed Point, $v^H$, exists and,
    \begin{align*}
    v < s_1(v) &, \ \forall v \in (\hat{v}, v^H) \\
    s_1(v) < v &, \ \forall v \in (v^H, 1)
    \end{align*}
\end{enumerate}
\end{proposition}

The proof is in Appendix \nameref{app:A2}. Recall from Prop. \ref{prop:5.5} that DI 2 corresponds to $\hat{v} < v < s_1(v)$, and D3 to $s_1(v) < v$.

\subsection{Gordon-Loeb Class II System Breach Probability Functions}

In this case, $D(s)=\frac{\xi(s)}{\alpha}$, so that $\hat{D}=\frac{1}{\alpha}\hat{\xi}$ and $\hat{s}=\hat{s}_\xi.$ (Recall that $\hat{\xi}\approx 0.64$ and $\hat{s}_\xi \approx 0.8$.) Since $s_1(\hat{v})=\hat{s}$, it follows that $s_1(\hat{v})=\hat{s}_\xi$. Also, since $D(\hat{s})=\frac{R}{f(\hat{v})} = -R\log \hat{v}$, it follows that $\hat{v}=\exp(\frac{-\hat{\xi}}{\alpha R})$.

The FPE in (\ref{eq:6.1}) translates to,
\begin{equation}\tag{6.3} 
\label{eq:6.3}
\xi(x) = -\alpha R \log(x), \ \ \hat{s} < x < 1
\end{equation}

\begin{proposition}
\label{prop:6.2}
For GL Class II functions,
\begin{enumerate}[label=(\roman*),itemsep=-3px]
    \item If $\alpha R < \frac{\hat{\xi}}{-\log \hat{s}_\xi} \approx 2.87$, then no Fixed Point exists, and,
    \begin{equation*}
        v < s_1(v), \ \forall v \in (\hat{v},1)
    \end{equation*}
    \item If $\frac{\hat{\xi}}{-\log \hat{s}_\xi} < \alpha R$, then one Fixed Point, $v^H$, exists, and,
    \begin{align*}
        s_1(v) < v &, \ \forall v \in (\hat{v}, v^H) \\
        v < s_1(v) &, \ \forall v \in (v^H,1)
    \end{align*}
\end{enumerate}
\end{proposition}
\textit{Proof:} It is easy to see that there exists one Fixed Point if and only if,
\begin{equation*}
    D(\hat{s}) < \frac{R}{f(\hat{s})}, \ \ \text{i.e.,} \ \ \frac{\hat{\xi}}{-\log \hat{s}_\xi} < \alpha R\ \ 
\end{equation*}
In the absence of any Fixed Point, the sign of $\{ s_1(v) - v\}$ is invariant for $v \in (\hat{v},1)$, and may be obtained from $v=\hat{v}$. If a Fixed Point $v^H$ exists, the sign of $\{ s_1(v) - v\}$ changes at $v^H$.

\subsection{General Results on the Existence of Fixed Points}

We give here a result that holds for $\gamma(s) \in \Gamma_1$, the set defined in Sec. \ref{sect:5.3}, which subsumes $\gamma(s)$ for GL Class I with $\gamma \leq 1$ and GL Class II. The result on the existence of solutions to (\ref{eq:6.1}) follows directly from the previously established properties of $D(x)$ and $f(x)$ for $x \in (\hat{s},1)$.

\begin{proposition}
Assume $\gamma(s) \in \Gamma_1$. If,
\begin{equation}\tag{6.4}
\label{eq:6.4}
    \frac{R}{f(\hat{s})} < \hat{D}
\end{equation}
then there exists a unique Fixed Point, unless $\frac{R}{f(1)}=0$, in which case no Fixed Point exists.
\end{proposition}
\textit{Proof:} Recall from Prop. \ref{prop:5.1} that $D(x)$ and $\frac{R}{f(x)}$ are strictly decreasing with increasing $x \in (\hat{s},1)$. Note that if (\ref{eq:6.4}) holds then $D(\hat{s}) > \frac{R}{f(\hat{s})}$, and since $D(1)=0 \leq \frac{R}{f(1)}$, it follows that a solution to (\ref{eq:6.1}) exists if the latter inequality is strict, and no solution exists if equality holds.

For GL Class II functions, $\frac{R}{f(1)}=0$, and hence from the above proposition, no Fixed Point exists if (\ref{eq:6.4}) holds, as also stated in Prop. \ref{prop:6.2}. When no Fixed Point exists, a consequence is that the Defender's optimal investment Decision Interval (DI) is invariant for all $v$, the initial vulnerability. Indeed, DI 1, i.e., where the investment decision is ``all in", is optimal.

In the following we draw attention to the role of $R$, the effective loss to gain ratio, in determining whether Fixed Points, i.e., solutions of (\ref{eq:6.1}) exist. Clearly no Fixed Point exists if $R$ is sufficiently large. Specifically, no fixed point exists if,
\begin{equation}\tag{6.5}
\label{eq:6.5}
    R > R_c
\end{equation}
where,
\begin{equation}\tag{6.6}
\label{eq:6.6}
    R_c = \max_{0 \leq x \leq 1} \left[ D(x) \hspace{0.1em} f(x) \right]
\end{equation}
The examples in Fig. \ref{fig:vary_R} show that for $R$ sufficiently large, no Fixed Point exists.

\subsection{Visualizing Fixed Point Equations}

\begin{figure}[ht]%
    \figuretag{6.1}
    \centering
    \captionsetup{justification=centering}
    \begin{subfigure}{6cm}
    \includegraphics[width=6cm]{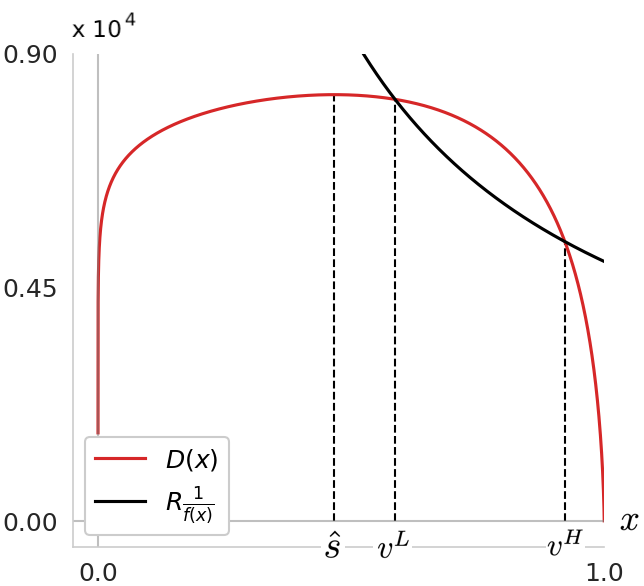}
    \caption{GL Class I: $R=5,000,\ \alpha=0.0001,\ \beta=1.1$,\\$\hat{s} \approx 0.47,\ v^L \approx 0.59,\ v^H \approx 0.92$}\label{fig:6.1.1}
    \end{subfigure}
    \qquad
    \begin{subfigure}{6cm}
    \includegraphics[width=\linewidth]{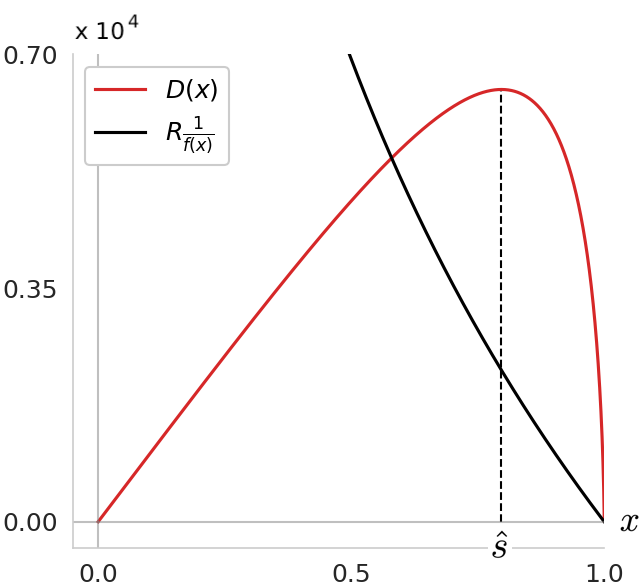}
    \caption{GL Class II:\\$R=10,000,\ \alpha=0.0001$}\label{fig:6.1.2}
    \end{subfigure}
    \caption{Visualizing the existence of Fixed Points in GL Class I and II functions.}
    \label{fig:vL_vH}%
\end{figure}

We now graphically illustrate various scenarios related to the existence of Fixed Points, and also provide corroboration of Prop. \ref{prop:6.1} and \ref{prop:6.2}. For the GL Class I function in Fig. \ref{fig:6.1.1}, $R=5,000$, $\alpha=0.0001$, and $\beta=1.1$, hence $\gamma \approx 0.9$. The gradient function $D(s)$ exhibits the inverted-U shape, in agreement with Prop. \ref{prop:6.1}, with the peak located at $\hat{s} \approx 0.47$. Also,
\begin{equation*}
    \xi(\hat{s})=\xi(0.47) \approx 0.43 \leq R \alpha \beta \approx 0.55 \leq \hat{\xi} \approx 0.64
\end{equation*}
and therefore, in correspondence to Prop. \ref{prop:6.1} (ii), we expect two Fixed Points, $v^L$ and $v^H$, to exist.

Fig. \ref{fig:6.1.2} displays a Class II function for which $R=10,000$ and $\alpha=0.0001$. In agreement with Prop. \ref{prop:6.2}, no Fixed Point solution to (\ref{eq:6.1}) exists, and, since $\gamma=0$ for all Class II functions, the gradient of $D(s)$ has the inverted-U shape, as stated in Prop. \ref{prop:5.1}.

\begin{figure}[ht]%
    \figuretag{6.2}
    \centering
    \captionsetup{justification=centering}
    \begin{subfigure}{6.5cm}
    \includegraphics[width=\linewidth]{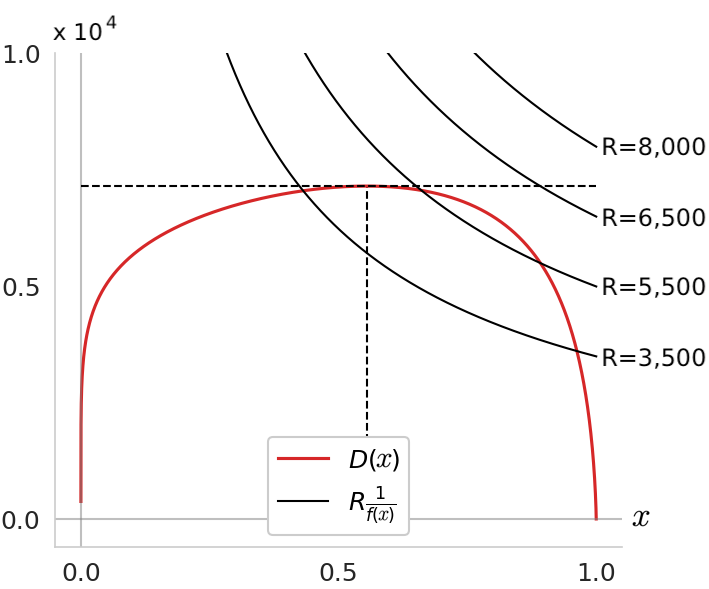}
    \caption{$\beta=1.2$, $(\gamma \approx 0.83)$}\label{fig:6.2.1}
    \end{subfigure}
    \qquad
    \begin{subfigure}{6.5cm}
    \includegraphics[width=6.5cm]{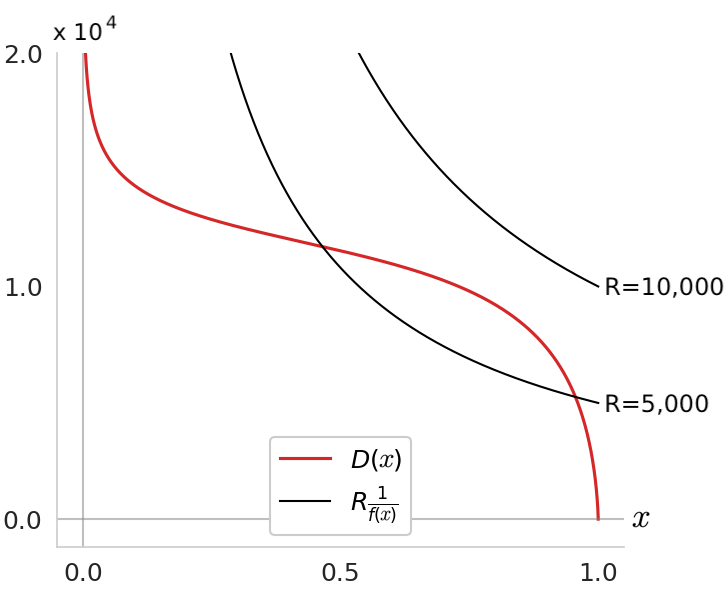}
    \caption{$\beta=0.9$, ($\gamma \approx 1.1$)}\label{fig:6.2.2}
    \end{subfigure}
    \caption{Modifying the parameter $R$ to capture Defender policies;\\GL Class I, $\alpha=0.0001$}%
    \label{fig:vary_R}%
\end{figure}

Fig. \ref{fig:vary_R} shows two GL Class I functions for which $\gamma(s)$ is constant, and $\gamma < 1$ and $\gamma > 1$, respectively, and the figure demonstrates the influence of the parameter $R$ in the determination of Fixed Points. Also observe in the example with $\gamma > 1$ that $D_s(s) < 0, \ \forall s \in (0,1)$, which corroborates the result in Sec. \ref{sect:5.3}.

\subsection{Attacker-Defender Model Examples}

Based on the understanding gained in the above examples of system breach probability functions, we proceed to obtain graphs of the optimal Attacker and Defender investments for various initial vulnerabilities, $v$. (Similar graphs are given in Gordon-Loeb \cite{ref:Gordon}.)

Consider the GL Class II function in Fig. \ref{fig:6.1.2} where $R=10,000$ and $\alpha=0.0001$. Since the Effective Loss to Gain Ratio, $R=\frac{L/d}{G/c}$, we let $L=\$100,000$, $G=\$100,000$, $d=\$1$, and $c=\$10,000$. That is, we assume that the Defender's potential loss is equal to the Attacker's potential gain, and, in the simplest representation, a single breach attempt by the Attacker costs $\$10,000$.

\begin{figure}[ht]
  \figuretag{6.3}
  \centering
  \captionsetup{justification=centering}
  \includegraphics[width=\linewidth]{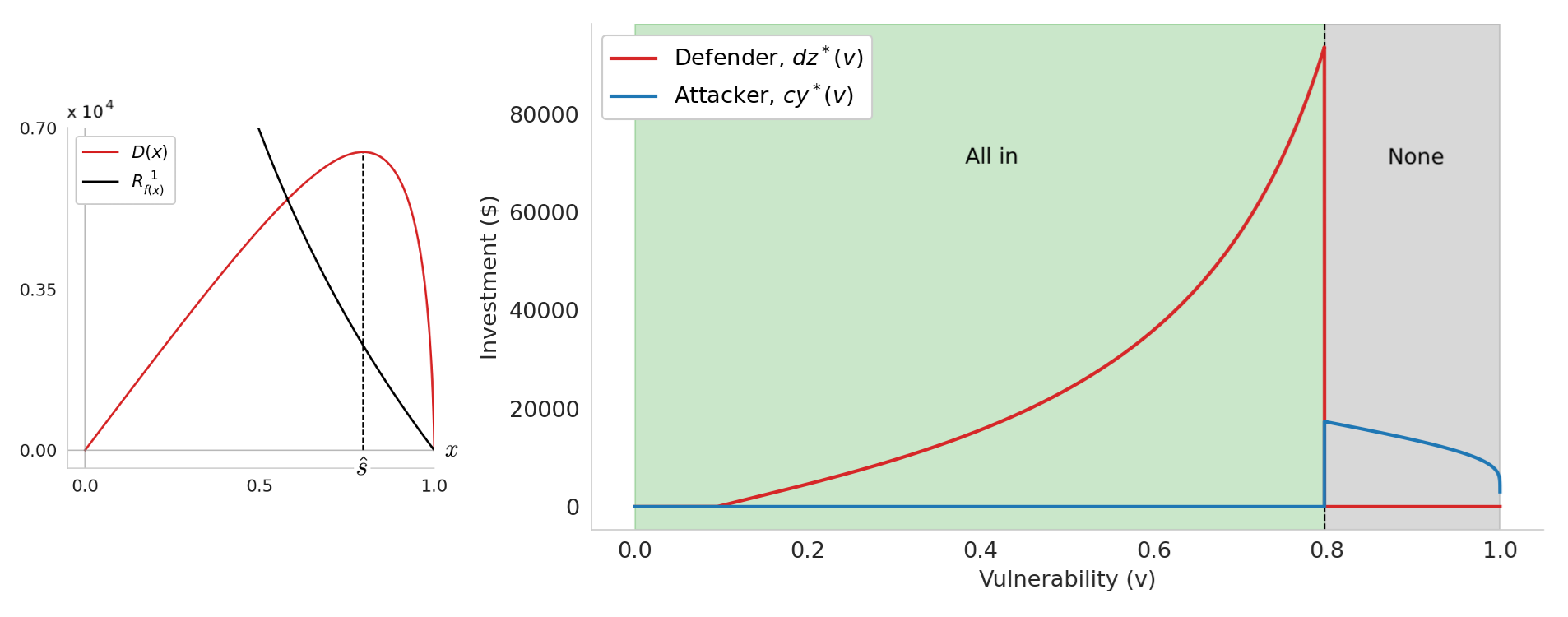}
  \caption{Example 1 -- Rational Attacker and Defender investments;\\GL Class II, $L=\$100,000$, $G=\$100,000$, $\alpha=0.0001$, $d=\$1$,  $c=\$10,000$}
  \label{fig:ex1}
\end{figure}

Since no solution to the FPE exists, the Defender's optimal decisions for all values of $v$ lie entirely in Decision Interval 2, i.e., it should choose to either pay the Price of Deterrence or invest nothing. This choice is made by comparing the value of $\Phi(s_P,v)$ and $\Phi(v,v)$. It is apparent from Fig. \ref{fig:ex1} that the transition from ``all in" to ``none" occurs at $v\approx 0.80$. The intuition behind this transition is two-fold:
\begin{enumerate}[label=(\roman*),itemsep=-3px]
    \item A rational Attacker will invest most heavily when the system vulnerabilities are in mid-range (see Fig. \ref{fig:y_s}). When the system vulnerability is high, the marginal benefit from investment is so low that further investment is not justified. Similarly, when the system is very invulnerable, the benefit of additional investment (as measured by the increase in expected gain), is small. A similar argument is made for Defenders in \cite{ref:Gordon}. Thus, there may be some interval of $s$ where the Defender's expected loss \textit{increases} with increasing effort since lowering the system breach likelihood may entice the Attacker to increase its effort.
    \item As $v$ increases, the Price of Deterrence, $z_P$, also increases. Thus, the Defender may find it impractical to fully deter an Attacker when the system vulnerability is high. Any smaller investment however, runs the risk of inducing large Attacker investments by reducing $s$ to be in mid-range.
\end{enumerate}

It is important to recognize that for $v < s_P$, $y^*(v) = z^*(v) = 0$, i.e., neither the Attacker nor the Defender finds it worthwhile to invest. We consider $s_P$ (and the ``all in" policy) to be the maximum vulnerability in $(0,v]$ that induces no Attacker investment, therefore, although no investment is being made by the Defender as well, we still consider this ``all in" since it is doing the bare minimum necessary to prevent any attack from occurring. 

Now, consider the GL Class I function from Fig \ref{fig:6.1.1} where $R=5,000$, $\alpha=0.0001$, and $\beta=1.1$. Specifically, let $L=\$100,000$, $G=\$70,000$, $d=\$1$, and $c=\$3,500$. In this case, the local minimum, $s_1(v)$, of the Defender's objective function plays an important role to induce an ``all or some" policy in certain intervals of $v$. As $v$ increases, the Defender's policies transition from ``all in", to ``all or some", then ``all or none", and finally back to ``all or some" at the $\hat{v}$, $v^L$, and $v^H$ interval boundaries, respectively. This example highlights the complexity of behavior that arises from our simple two-sided model.

\begin{figure}[ht]
  \figuretag{6.4}
  \centering
  \captionsetup{justification=centering}
  \includegraphics[width=\linewidth]{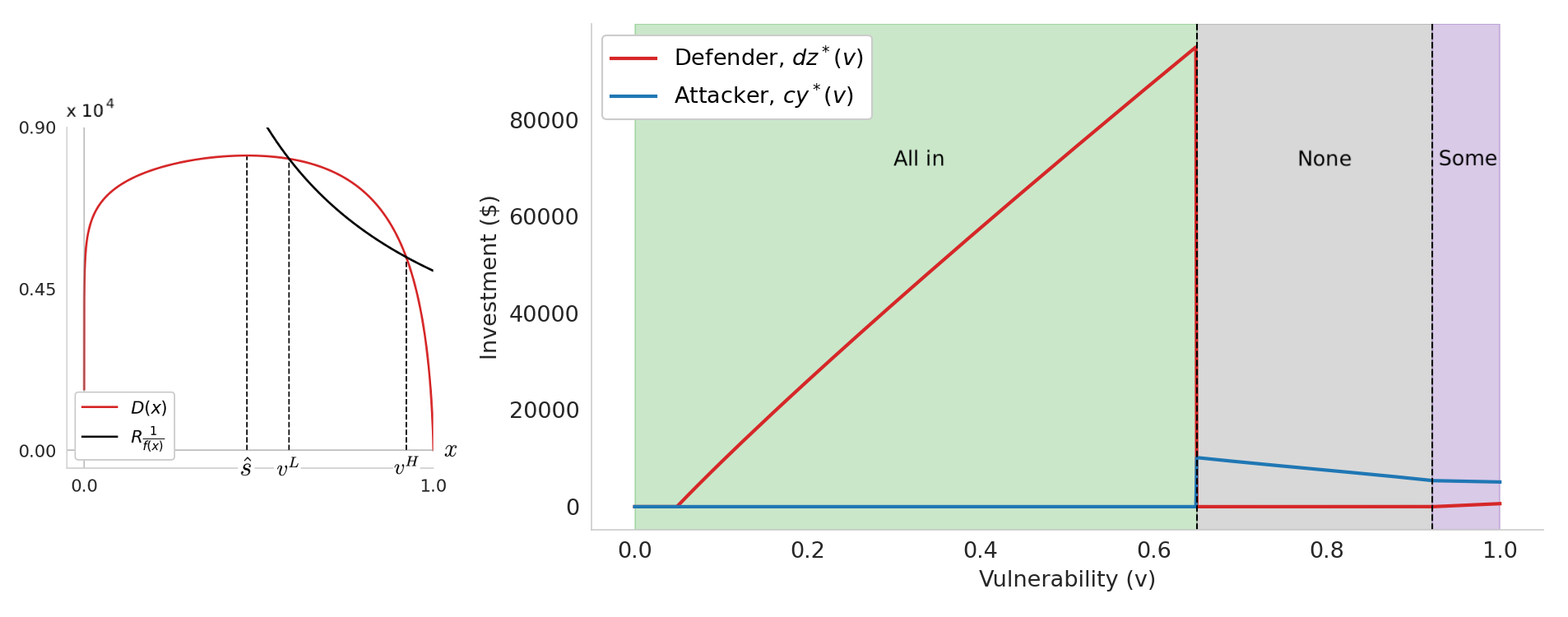}
  \caption{Example 2 -- Rational Attacker and Defender investments;\\GL Class I, $L=\$100,000$, $G=\$70,000$, $\alpha=0.0001$, $\beta=1.1$, $d=\$1$, and $c=\$3,500$\\$\hat{s} \approx 0.47,\ v^L \approx 0.59,\ v^H \approx 0.92$}
  \label{fig:ex2}
\end{figure}

\subsection{Comparisons with the Gordon-Loeb Model}

There are several interesting features worth highlighting in Fig. \ref{fig:ex1} and \ref{fig:ex2}. In contrast to the results from the model of Gordon-Loeb \cite{ref:Gordon}, optimal investment curves here are not guaranteed to be smooth. This reflects the stark transitions in decision type, e.g., from ``all in" to ``none", as the initial system vulnerability is varied.

\begin{figure}[ht]
  \figuretag{6.5}
  \centering
  \captionsetup{justification=centering}
  \includegraphics[width=13cm]{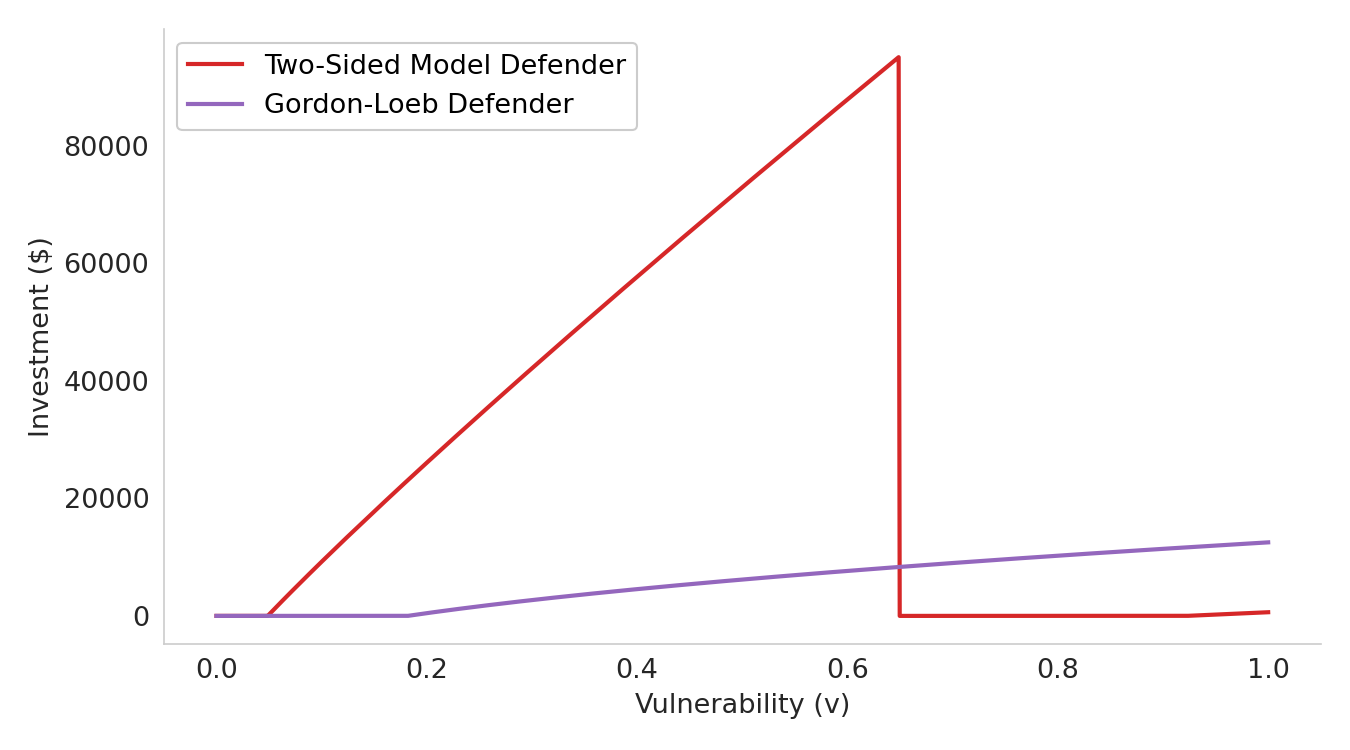}
  \caption{Comparison of Defender investments from the Gordon-Loeb and our two-sided models.\\GL Class I: $L=\$100,000$, $G=\$70,000$, $\alpha=0.0001$, $\beta=1.1$, $d=\$1$, $c=\$3,500$}
  \label{fig:ex3}
\end{figure}

Furthermore, we see a substantial departure from the celebrated $1/e$ result \cite{ref:Gordon}. In fact, the results from our model indicate that a rational, risk-neutral Defender may find it in its best interest to invest up to the potential loss of $L$ due to a system breach.\footnote{Of note is that the upper limit to rational investment is $L \hspace{0.1em} T(S(0,v))$, however $T(S(0,v))$ may tend arbitrarily close to $1$.} For an appreciation of these differences, compare the optimal Defender investment curve from Fig. \ref{fig:ex2} for our Attacker-Defender model with that from the Gordon-Loeb model, which are shown in Fig. \ref{fig:ex3}.

Observe that the inclusion of a rational, risk-neutral Attacker has a profound impact on the Defender's investment strategy. We find that, in general, for broad ranges of the system's initial vulnerability the optimal investments from the Gordon-Loeb model \cite{ref:Gordon} underestimate what is necessary to prudently defend the system. This is because any attack, regardless of severity, serves to increase system breach likelihood (and therefore increase the Defender's expected loss) by some degree. Thus, we argue that it is crucial to consider the goals and resources of an Attacker so as to capture this far-reaching characteristic of cybersecurity and arrive at more realistic estimates of investments in system security.

\section{Conclusions}
\label{sect7}

We conclude with some observations on directions for future work. We have noted earlier that the estimates of potential gain ($G$) and loss ($L$) play an important role, via the parameter $R$, in the determination of the optimal investment policy. It follows that a non-standard dimension of cyberwarfare is the promotion of false estimates by adversaries, and a better understanding of how to deal with this possibility will be useful. We have assumed a risk-neutral Defender, and a question of interest, albeit one that will introduce added complexity, is the nature of the optimal investment for a risk-averse Defender. In a similar vein, it will be insightful to obtain the Defender's decision based on worst-case analysis to allow for a possibly irrational Attacker.

The deployment costs of attacks and defense have been modeled as linear functions of the respective efforts. Realistically the costs may be expected to be nonlinear, and to also include large fixed costs, which are known to significantly affect market structures and pricing \cite{ref:Panzar1, ref:Braeutigam}. It should be worthwhile to delineate areas in cybersecurity where fixed costs matter and where they do not.

A natural progression of our static problem formulation would be to dynamic, sequential Stackelberg games. Much is known \cite{ref:Breton}, but not where the focus is on fundamental understanding in the context of the economics of cybersecurity. Other avenues for generalizing our model is to networked systems, where attack graphs provide a possible framework \cite{ref:Wang}. Similarly, Moving Target Defense \cite{ref:Lei} is also a promising direction for generalization.

\section*{Appendix}

\subsection*{A.1}
\label{app:A1}

\textit{Proof of Prop. \ref{prop:5.1}}: Taking the derivative with respect to $s$,
\begin{equation}\tag{A1.1}
\label{eq:A1.1}
    s^2 H_s(s;\gamma) = \frac{-2 \left\{ s^2 - (1-s) \log^2 (1-s) \right\}}{(1-s) \log^2(1-s)} - \left\{ 1 - \gamma(s) \right\} - \gamma_s(s) s (1-s)
\end{equation}

Now denoting the numerator of the first term on the right hand side (rhs) by $-2 F(s)$, it can be shown that,
\begin{align}\tag{A1.2}
\label{eq:A1.2}
    F(s) = s^2 - (1-s) \log^2(1-s) &> 0, \ \ s \in (0,1), \\[0px]
    &=0, \ \ s = 0. \nonumber
\end{align}
To see this, observe that $F(s)$ is strictly convex since,
\begin{equation*}
    \frac{1}{2}(1-s)F_{ss}(s) = - \left\{ s + \log(1-s) \right\} > 0, \ \ s \in (0,1)
\end{equation*}
Also, $F_s(0) = 0$, hence $F_s(s) > 0, \ s \in (0,1)$. Since $F(0) = 0$, (\ref{eq:A1.2}) follows.

Using (\ref{eq:A1.2}) and (\ref{eq:A1.1}), it follows that if $0 \leq \gamma(s) < 1$ and $0 \leq \gamma_s(s), \ \forall s \in [\hspace{0.1em}0,1]$, then $H_s(s;\gamma) < 0, \ s\in [\hspace{0.1em}0,1]$. Also, $0 \leq H(0;\gamma)$ and $H(1;\gamma) < 0$. Hence there exists a unique $\hat{s} \in (0,1)$ such that $H(s;\gamma)=0$, i.e., $D_s(\hat{s})=0$ and $D_s(s) > (<) \ 0$ for $s < (>) \ \hat{s}$. 

\vspace{15px}

\textit{Proof of Prop. \ref{prop:5.2}}: For $1 < \gamma(s), \ s \in [\hspace{0.1em}0,1]$, from (\ref{eq:4.7}),
\begin{align}
\left[ -s \log(1-s) \right] H(s;\gamma) &= 2s + (2-s)\log(1-s) + \left\{ \gamma(s)-1 \right\}(1-s)\log(1-s) \label{eq:A1.4} \tag{A1.4} \\
           &\leq 2s + (2-s)\log(1-s) \label{eq:A1.5} \tag{A1.5} \\
           &< 0, \ \ s \in (0,1] \label{eq:A1.6} \tag{A1.6}
\end{align}

To see how (\ref{eq:A1.6}) follows from (\ref{eq:A1.5}), let $h(s)$ denote the expression on the rhs of (\ref{eq:A1.5}), and verify that $h_{ss} < 0$, i.e., $h(s)$ is strictly concave in $(0,1]$. Also, $h_s(0)=0$ and $h(0)=0$, which in combination with strict concavity yields (\ref{eq:A1.6}).

Since the bracketed term in the lhs in (\ref{eq:A1.4}) is positive for $s \in (0,1]$, we have $H(s;\gamma) < 0$ and $D_s(s) < 0$, $s \in (0,1]$. In the definition of $D(s)$ in (\ref{eq:5.2.2}), note that as $s \rightarrow 0$, $\log^2(1-s) \sim s^2$, and, from (\ref{eq:2.18})-(\ref{eq:2.19}), $g(s)=O(s^\delta)$. Hence, as $s \rightarrow 0$,
\begin{equation}\tag{A1.7}
\label{eq:A1.7}
    D(s) \rightarrow O(s^{2-\delta})
\end{equation}
From (\ref{eq:2.19}), when $\gamma(0) > 1$, $\delta > 2$, and consequently $D(s) \rightarrow \infty$ as $s \rightarrow 0$.

\subsection*{A.2}
\label{app:A2}

\textit{Proof of Prop. \ref{prop:6.1}}: First we show that,
\begin{equation} \tag{A2.1}
\label{eq:A2.1}
    s_1(\hat{v}) = \hat{s} < \hat{v}
\end{equation}
To see this, note from the defining relation $D(\hat{s}) = \frac{R}{f(\hat{v})}$ that,
\begin{equation} \tag{A2.2}
\label{eq:A2.2}
    \xi(\hat{s}) = R \alpha \beta \left( \frac{\hat{s}}{\hat{v}}\right)^{1/\beta}
\end{equation}
since $D(s)= \xi(s) / (\alpha \beta s^{1/\beta})$. Now suppose, contrary to (\ref{eq:A2.1}), $\hat{s} \geq \hat{v}$. Then from (\ref{eq:A2.2}), $\xi(s) \geq R \alpha \beta$, which contradicts $\hat{\xi} < R \alpha \beta$.

If $\hat{\xi} < R \alpha \beta$, then from (\ref{eq:6.2.2}), clearly no Fixed Point exists. Now noting (\ref{eq:A2.1}), in the absence of any Fixed Point, it must be that $s_1(v) < v, \ \forall v \in (\hat{v},1)$. Hence (i) is proved.

Next we show that,
\begin{equation}\tag{A2.3}
\label{eq:A2.3}
    \hat{s} < \hat{s}_\xi \ \ \text{and} \ \ \xi(\hat{s}) < \hat{\xi}
\end{equation}
From the relation between $D(s)$ and $\xi(s)$ given above,
\begin{equation}\tag{A2.4}
\label{eq:A2.4}
    \alpha \beta D_s(s) = \frac{1}{s^{1/\beta}} \left[ \xi_s(s) - \frac{1}{s}\xi(s) \right].
\end{equation}
Since $D_s(\hat{s})=0$, it follows that,
\begin{equation}\tag{A2.5}
\label{eq:A2.5}
    \xi_s(\hat{s}) = \frac{1}{\hat{s}}\xi(\hat{s}) > 0
\end{equation}
Hence, from the inverted-U shape of $\xi(s)$ and the fact that its maximum value of $\hat{\xi}$ is when $s=\hat{s}_\xi$, (\ref{eq:A2.3}) follows.

If $R \alpha \beta < \hat{\xi}$, then from the inverted-U shape of $\xi(s)$, it follows that there exist two solutions, $\xi_1$ and $\xi_2$, to the solution $\xi(s)=R\alpha \beta$, and we let $\xi_2 < \hat{s}_\xi < \xi_1$. Making use of (\ref{eq:A2.3}), the possibilities are,
\begin{equation}\tag{A2.6}
\label{eq:A2.6}
    \hat{s} < \xi_2 < \hat{s}_\xi < \xi_1, \ \ \text{and} \ \ \xi_2 < \hat{s} < \hat{s}_\xi < \xi_1.
\end{equation}
In the former case, two Fixed Points exist, $v^L = \xi_2$ and $v^H=\xi_1$, and in the latter case only one Fixed Point exists, $v^H=\xi_1$. (In the latter case, since $\xi_2 < \hat{s},\ \xi_2$ violates the constraint that Fixed Points satisfy.) We can precisely characterize the separation of the two cases in (\ref{eq:A2.6}). The two cases correspond respectively to,
\begin{equation}\tag{A2.7}
\label{eq:A2.7}
    \xi(\hat{s}) < \xi(\xi_2)=R \alpha \beta < \hat{\xi}, \ \ \text{and} \ \ \xi(\xi_2)=R\alpha \beta < \xi(\hat{s}) < \hat{\xi}.
\end{equation}
The former case corresponds to (ii) of the Proposition, and the latter to (iii).

In both cases (ii) and (iii), the sign of $\{s_1(v)-v\}$ between the transition points follows from the number of Fixed Points in each case, in combination with the sign of $\{s_1(\hat{v})-\hat{v}\}$, which has been established in (\ref{eq:A2.1}).

\section*{Acknowledgements}

DM gratefully acknowledges the support of the National Institute of Standards and Technology, and, especially, NIST’s Applied and Computational Mathematics Division. DM also gratefully acknowledges the benefit of discussions with Vladimir Marbukh.

\vspace{5px}

\newpage

\setstretch{1}
\footnotesize{
	\bibliographystyle{abbrv}
	\bibliography{bibliography}

\begin{thebibliography}{100}

\vspace{5px}

\bibitem[1]{ref:Clark} ``At the Nexus of Cybersecurity and Public Policy: Some Basic Concepts and Issues", D. Clark, T. Berson, H.S. Lin, Editors, National Academies Press, Washington DC; 2014.

\bibitem[2]{ref:Smith} Z. M. Smith and E. Lostri, ``The Hidden Costs of Cybercrime", Report by the Center for Strategic and International Studies and McAfee, \url{https://www.mcafee.com/enterprise/en-us/assets/reports/rp-hidden-costs-of-cybercrime.pdf}

\bibitem[3]{ref:IBM} ``Cost of a Data Breach Report 2021", IBM report
\url{https://www.ibm.com/downloads/cas/OJDVQGRY}

\bibitem[4]{ref:NVD} National Vulnerability Database, \url{https://en.wikipedia.org/wiki/National_Vulnerability_Database}

\bibitem[5]{ref:Mell} P. Mell, K. Scarfone, S. Romanosky, ``A Complete Guide to the Common Vulnerability Scoring System", in NIST CVSS, National Institute of Standards and Technology, June 2007.

\bibitem[6]{ref:Danskin} Danskin, J.M., ``The Theory of Max-Min and Its Application to Weapons Allocation Problems", Springer-Verlag, NY, 1967.

\bibitem[7]{ref:Gordon} Gordon, L. and Loeb, M. (2002), ``The Economics of Information Security Investment", ACM Trans. on Information and System Security, 5 (4): 438-457.

\bibitem[8]{ref:Baryshnikov} Baryshnikov, Y. (2012) ``IT Security Investment and Gordon-Loeb's 1/e Rule", Proceedings of the 11th Workshop on the Economics of Information Security (WEIS), Berlin, 25-26 June 2012.

\bibitem[9]{ref:Lelarge} Lelarge, M. (2012) ``Coordination in Network Security Games: A Monotone Comparative Statics Approach", IEEE Journal on Selected Areas in Communications, 30 (11): 2210-2219.

\bibitem[10]{ref:Gibbons} R. Gibbons, ``Game Theory for Applied Economists", Princeton University Press, 1992.

\bibitem[11]{ref:Rockafellar} R.T. Rockafellar, ``Lagrange multipliers and optimality", SIAM Review 35(2),1993, 183-238.

\bibitem[12]{ref:Anderson1} R. Anderson, ``Why information security is hard -- An economic perspective", in Proceedings of 17$^{\text{th}}$ Annual Computer Security Applications Conference (ACSAC), New Orleans, Dec. 2001.

\bibitem[13]{ref:Anderson2} R. Anderson, C. Barton, R. Bohme, R. Clayton, M.J.G. van Eeten, M. Levi, T. Moore, S. Savage, ``Measuring the Cost of Cybercrime", in Proceedings of 11th Workshop on the Economics of Information Security, June 2012.

\bibitem[14]{ref:Tiwari} R.K.Tiwari, K. Karlapalem, ``Cost Tradeoffs for Information Security Assurance", in Proceedings of 4th Annual Workshop on the Economics of Information Security, WEIS, Harvard University, June 2005.

\bibitem[15]{ref:Tanaka} H. Tanaka, K. Matsuura, O. Sudoh, ``Vulnerability and information security investment: An empirical analysis of e-local government in Japan", Journal of Accounting and Public Policy, vol. 24, 2005, p 35-59. 

\bibitem[16]{ref:Mazurczyk} W. Mazurczyk, L. Caviglione, ``Cyber Reconnaissance Techniques", Communications of the ACM, March 2021, vol. 54, no. 3, pp 86-95.

\bibitem[17]{ref:Wang} L. Wang, T. Islam, T. Long, A. Singhal, S. Jajodia, ``An Attack Graph Based Probabilistic Security Metric", in Proceedings 22$^{\text{nd}}$ IFIP WG 11.3 Working conference on Data and Application Security, London, UK, July 2008.

\bibitem[18]{ref:Lei} C. Lei, H.-Q. Zhang, J.-L. Tan, Y.-C. Zhang, and X.-H. Liu, ``Moving Target Defense Techniques: A Survey", Security and Communication Networks, Volume 2018, Article ID 3759626.

\bibitem[19]{ref:Breton} M. Breton, A.Ali, A. Haurie, ``Sequential Stackelberg Equilibria in Two-Person Games", Journal of Optimization Theory and Applications, vol. 59, no.1, Oct. 1988, p71- 97.

\bibitem[20]{ref:Gueye} A. Gueye, V. Marbukh, ``A Game-Theoretic Framework for Network Security Vulnerability Assessment and Mitigation", in J. Grossklags and J.Walrand  Editors, GameSec 2012, LNCS 7638, 2012, pp 186-200.

\bibitem[21]{ref:Baumol} Baumol, W.J., Panzar, J.C. and Willig, R.D. (1982) ``Contestable markets and the theory of industry structure". New York: Harcourt Brace Jovanovich.

\bibitem[22]{ref:Panzar1} Panzar, J.C., (1989), ``Technological Determinants of Firm and Industry Structure", Chapter 1, Handbook of Industrial Organization, Volume II, Ed. R. Schmalensee and R.D. Willig, Elsevier Science Publishers B.V.

\bibitem[23]{ref:Braeutigam} Braeutigam, R.R., (1989), ``Optimal Policies for Natural Monopolies", Chapter 23, Handbook of Industrial Organization, Volume II, Ed. R. Schmalensee and R.D. Willig, Elsevier Science Publishers B.V.

\bibitem[24]{ref:Panzar2} Panzar, J.C. and Willig R.D. (1977) ``Free entry and sustainability of natural monopoly", Bell Journal of Economics, 8:1-22.

\bibitem[25]{ref:Faulhaber} Faulhaber, G.R. (1975) ``Cross-subsidization: Pricing in public enterprises", American Economic Review, 65:966-977.

\bibitem[26]{ref:Swift} Swift, J, ``Soviet-American Arms Race", History Today, Published in History Review Issue 63 March 2009 \url{https://www.historytoday.com/archive/soviet-american-arms-race}

\bibitem[27]{ref:Goodman} Goodman, I, ``The Price of Victory in Cold War is \$5.8 Trillion for Nuclear Arms and Delivery Systems, Says Panel”, Physics Today, 51 (8), Aug. 1998, p 49.

\bibitem[28]{ref:Aghion} ``Competition and Innovation: An Inverted-U Relationship", P. Aghion et. al., The Quarterly J. of Econ., vol 120, no.2, May 2005, pp 701--728.

\end{thebibliography}
}

\end{document}